\documentclass[aps,prb,reprint,groupedaddress]{revtex4-1}
\usepackage{graphicx}
\usepackage{gensymb}
\usepackage{amsmath}
\usepackage[colorlinks,hyperindex]{hyperref}
\hypersetup
{
colorlinks,%
citecolor=blue,%
linkcolor=blue,%
urlcolor=blue,%
}

\begin{document}

\title{Selection Rules for Quasi-Bound States in the Continuum}


\author{Adam C. \surname{Overvig}}
\author{Stephanie C. \surname{Malek}}
\author{Michael J. \surname{Carter}}
\author{Sajan \surname{Shrestha}}
\author{Nanfang \surname{Yu}}
\email[Correspondence to: ]{ny2214@columbia.edu}
\affiliation{Department of Applied Physics and Applied Mathematics, Columbia University, New York, NY 10027}

\date{\today}

\begin{abstract}
Photonic crystal slabs (PCSs) are a well-studied class of devices known to support optical Fano resonances for light normally incident to the slab, useful for narrowband filters, modulators, and nonlinear photonic devices. In shallow-etched PCSs the linewidth of the resonances is easily controlled by tuning the etching depth. This design strength comes at the cost of large device footprint due to the poor in-plane localization of optical energy. In fully-etched PCSs realized in high index contrast material systems, the in-plane localization is greatly improved, but the command over linewidth suffers. This disadvantage in fully-etched PCSs, also known as high contrast gratings (HCGs), can be overcome by accessing symmetry-protected Bound States in the Continuum (BICs). By perturbing an HCG, the BIC may be excited from the free space with quality factor showing an inverse squared dependence on the magnitude of the perturbation, while inheriting the excellent in-plane localization of their unperturbed counterparts. Here, we report an exhaustive catalog of the selection rules (if and to which free space polarization coupling occurs) of symmetry-protected BICs controlled by in-plane symmetry breaking in six types of two-dimensional PCS lattices. The chosen lattices allow access to the three highest symmetry mode classes of unperturbed square and hexagonal PCSs. The restriction to in-plane symmetry breaking allows for manufacturing devices with simple lithographic fabrication techniques in comparison to out-of-plane symmetry breaking, useful for practical applications. The approach reported provides a high-level roadmap for designing PCSs supporting controllable sharp spectral features with minimal device footprints using a mature fabrication platform. To demonstrate the use of the resulting alphabet of structures, we numerically demonstrate nonlocal metasurface platforms for Terahertz generation,  mechnically tunable optical lifetimes, and wavefront shaping exclusively at resonance. 
\end{abstract}

\maketitle

\section{Introduction}
Enhancement of light-matter interactions is a key capability for improving and expanding the functionality of a wide gamut of photonic devices. Spatially and temporally confining light enables compact planar optical modulators with fast switching speeds~\cite{xu_12.5_2007,nguyen_compact_2012,melikyan_high-speed_2014,gao_electro-optic_2014}, narrowband bandpass filters~\cite{wang_guided-mode_1990,wang_theory_1993,tibuleac_reflection_1997,rosenblatt_resonant_1997,cui_normal_2015}, sensitive biological and refractive index sensors~\cite{zhuo_label-free_2015,scullion_slotted_2011,tittl_imaging-based_2018}, efficient optical micro-electromechanical devices~\cite{antoni_deformable_2011,kemiktarak_optomechanics_2014}, novel lasers~\cite{bigelow_observation_2003,zhang_visible_2006,noda_photonic_2010,chase_1550_2010, nguyen_directional_2018}, and enhanced nonlinear~\cite{corcoran_green_2009,galli_low-power_2010,lin_cavity-enhanced_2016,lee_giant_2014} and quantum optical phenomena~\cite{yoshie_vacuum_2004,wang_quantum_2018}. This is conventionally achieved by the introduction of an optical cavity, which circulates optical energy, affording a photon many passes through a material.

Planar diffractive optics enable uniquely compact optical confinement in lightweight quasi-two-dimensional systems fabricated by mature micro- and nano-fabrication technologies. Traditional plasmonic materials such as gold enable strong light-matter interaction in metasurfaces~\cite{yu_light_2011,lukyanchuk_fano_2010,cui_dynamic_2012,khanikaev_fano-resonant_2013,li_graphene_2015,lee_giant_2014}, but are incompatible with standard complementary metal-oxide semiconductor (CMOS) foundries. Alternative plasmonic materials are an active area of study\cite{naik_alternative_2013,wang_tunability_2017}, but without exception introduce substantial optical losses that reduce the efficiency of a photonic device. These limitations motivate exploring methods of confining optical energy without metals, restricting the optical materials to common dielectric materials such as silicon and its oxide.

A classic example of a dielectric diffractive optical element with enhanced light-matter interactions is the low-contrast grating (LCG), or guided mode resonance filter~\cite{wang_guided-mode_1990,wang_theory_1993,tibuleac_reflection_1997,rosenblatt_resonant_1997}. By periodically corrugating a thin slab with subwavelength periodicity, a laterally propagating waveguide mode supported by the slab may couple to normally incident light. The leakage out of the slab interferes with the direct optical pathways (here, the Fabry-Perot resonance), producing a well-known Fano resonance~\cite{fano_sullo_1935,fano_effects_1961,limonov_fano_2017}. Related phenomena have been studied for over a century, beginning with Wood’s anomalies~\cite{wood_xlii._1902,hessel_new_1965,enoch_theory_2012}. In an LCG, the degree of corrugation can be easily controlled experimentally, and is a design parameter that directly controls the linewidth of the resonant spectral feature. In particular, for small corrugation the Q-factor of the resonance is known to be inversely related to the depth of the corrugation~\cite{fan_temporal_2003}. However, this attractive design feature comes with an inherent drawback: the long optical lifetime comes from the long distance the guided mode travels within the device before coupling back to free space; the device therefore needs to be of a lateral size comparable to this characteristic travel distance in order to observe a narrow spectral feature. In other words, LCGs are constrained by a tradeoff between spatial confinement (device size) and temporal confinement (Q-factor).

Another well-studied diffractive optical element is the high contrast grating (HCG)~\cite{chang-hasnain_high-contrast_2012,karagodsky_theoretical_2010,chang-hasnain_high-contrast_2011}, known to enable compact devices due to large in-plane Bragg reflection laterally confining optical energy. Since the corrugation is deep (and, typically, complete) in HCGs, the ease of control of the Q-factor by the method present in LCGs is lost. HCGs are best-known for their broadband spectral features for this reason. However, HCGs are also known to support sharp spectral features in the form of Fano resonances~\cite{chang-hasnain_high-contrast_2011,yoon_critical_2016,wang_optical_2016}. In particular, for certain combinations of optical materials, geometries, wavelength, angle, and polarization, the Q-factor may become infinite, a phenomenon known as a ``bound state in the continuum'' (BIC)~\cite{plotnik_experimental_2011,hsu_observation_2013,bulgakov_bloch_2014,hsu_bound_2016}. Operating near a BIC in the relevant multi-dimensional parameter space allows tuning of a resonance with finite Q-factor. Unfortunately, because of the complex and sensitive dependence on many parameters simultaneously, this control is not robust in comparison to the control in an LCG.

However, HCGs can support two classes of BICs: those excluded from coupling to free space due to symmetry constraints (or ``symmetry-protected''), and those excluded for reasons unrelated to symmetry (or ``accidental''~\cite{sadrieva_multipolar_2019}). It has been argued recently~\cite{overvig_dimerized_2018} that symmetry-protected BICs in HCGs are better suited than accidental BICs for creating compact optical devices with sharp spectral features. It is well-known that by reducing the symmetry~\cite{lemarchand_study_1999,Bingham_08,zeng_tunable_2015,nguyen_symmetry_2017, liang_symmetry-reduced_2014,cui_normal_2016,qiu_active_2012,cui_dynamic_2012,kilic_controlling_2008,foley_symmetry-protected_2014,fedotov_sharp_2007,nguyen_symmetry_2018,cueff_tailoring_2018} of an HCG or PCS, symmetry-protected BICs become quasi-bound in the continuum, at which point they are referred to as ``quasi-BICs"\cite{koshelev_asymmetric_2018}. Quasi-BICs couple to light at normal incidence with optical lifetimes controlled by the magnitude of the perturbation that breaks the symmetry protecting them, thereby restoring a robust design paradigm for controlling the Q-factor of a sharp spectral feature. Furthermore, it has also recently been shown~\cite{nguyen_symmetry_2018,cueff_tailoring_2018} that proper perturbation (including breaking vertical symmetry) allows excellent control of the band structure. Therefore, a symmetry-broken HCG inherits the benefits of both LCGs and HCGs relevant to sharp spectral features in compact devices.

In particular, a period doubling perturbation (a dimerization of an HCG) allows modes previously bound (under the light line at the edge of the first Brillouin zone (FBZ)) to be brought into the continuum, coupling to a range of angles near normal incidence to a degree controlled solely by the perturbation~\cite{wu_spectrally_2014,liang_symmetry-reduced_2014,zeng_tunable_2015,lan_dark_2017,nguyen_symmetry_2017,qiu_active_2012,nguyen_symmetry_2018,cueff_tailoring_2018}. Consequently, a ``dimerized high contrast grating'' (DHCG~\cite{overvig_dimerized_2018}) is an excellent candidate platform for planar optical devices with both spatial and temporal confinement of light. Much of the study of DHCGs has focused on simple, one-dimensional devices, enabling control of the mode in one in-plane direction, but not in the orthogonal direction. Two-dimensional, high-index contrast PCSs with periodic perturbations are the natural extension of DHCGs that solve this limitation, and are the subject of this paper. The number of symmetries in a two-dimensional PCS is significantly greater than the simple one-dimensional case; the wealth of modal interactions between free space and two-dimensional PCSs with periodic perturbation therefore requires detailed exploration.

Symmetry-protected BICs are commonly studied in monatomic PCSs, where even/odd symmetry conditions may preclude coupling to free space at normal incidence. The BICs in diatomic PCSs (e.g., DHCGs) are subject to the analogous even/odd symmetry conditions, so that once the period doubling has folded the bound modes into the continuum, they may still be left bound in the continuum. If the relevant even/odd symmetry is broken, the symmetry-protected BICs become quasi-BICs. Both the monatomic and diatomic approaches fall under the same category of symmetry-protected BICs, but access distinct high symmetry modes (that is, modes with distinct periodicity and field profiles). Therefore, to fully utilize the available perturbations and modes, we study both monatomic and multi-atomic PCSs. 

We note that the simplest method of breaking the relevant in-plane symmetries is to excite quasi-BICs with light at an incident angle just off the substrate's normal. However, as argued in ~\cite{overvig_dimerized_2018}, this approach is restrictive in comparison to breaking the symmetry by perturbing the structure. In particular, this method works only for light with a limited angular spread of optical momentum centered at a precise angle (to get the target Q-factor), therefore requiring devices with large footprints and precise tilt of the device relative to the source optics.

Lastly, while in this Article we describe quasi-BICs as supported by all-dielectric PCSs, we note that the Group Theory approach taken here is valid irrespective of materials system, so long as the materials are isotropic. For instance, arrays of Silicon pillars are treated the same as arrays of holes in a Silicon (or even metal) slab. The choice of materials system may be made based on the needs of the application; an analogous set of modes (as classified by their in-plane symmetries) obeying the selection rules derived here will exist, albeit with widely varying field profiles and resonant frequencies. 

In this paper, we study the optical response due to in-plane perturbations applied to high-symmetry PCS lattices. Throughout, we assume these PCSs have cross-sections invariant in the out-of-plane direction, and we assume any out-of-plane symmetry introduced by the presence substrate is negligible. In Sec.~\ref{Back}, we review quasi-BICs and how they spatially and temporally confine light. In Section~\ref{B2}, we review the classification in the language of Group Theory of the three types of high symmetry modes supported by each of square and hexagonal PCSs (six types of modes in total). Section~\ref{S1} explores six classes of perturbed lattices chosen to target the six high symmetry modes. To determine the impact of these perturbations, Sec.~\ref{S2} derives the symmetry constrained coupling conditions specifying which, if any, free space polarization couples upon perturbation. The degenerated space groups compatible with each unperturbed lattice are exhaustively listed, and in Sec.~\ref{S3} the polarization dependence for each mode and each space group is written down by applying Group Theory principles. The result is a catalog of the selection rules for quasi-BICs describing all the unique ways that the six highest symmetry modes of square and hexagonal photonic crystal lattices may be accessed by in-plane symmetry breaking. Finally, in Sec.~\ref{D} we discuss notable aspects of the catalog and sketch several example device applications. In particular, here and in the accompanying Letter~\cite{overvig_multifunctional_2020} we show how the catalog guides device design using successive perturbations to achieve multifunctional control of optical spectra, and introduce a novel class of metasurfaces that use this multifunctional control to spatially control resonant wavefronts.

\section{Background}
\label{Background}

\subsection{Quasi-Bound States in the Continuum}
\label{Back}

\begin{figure*}
\includegraphics[width=1.8\columnwidth]{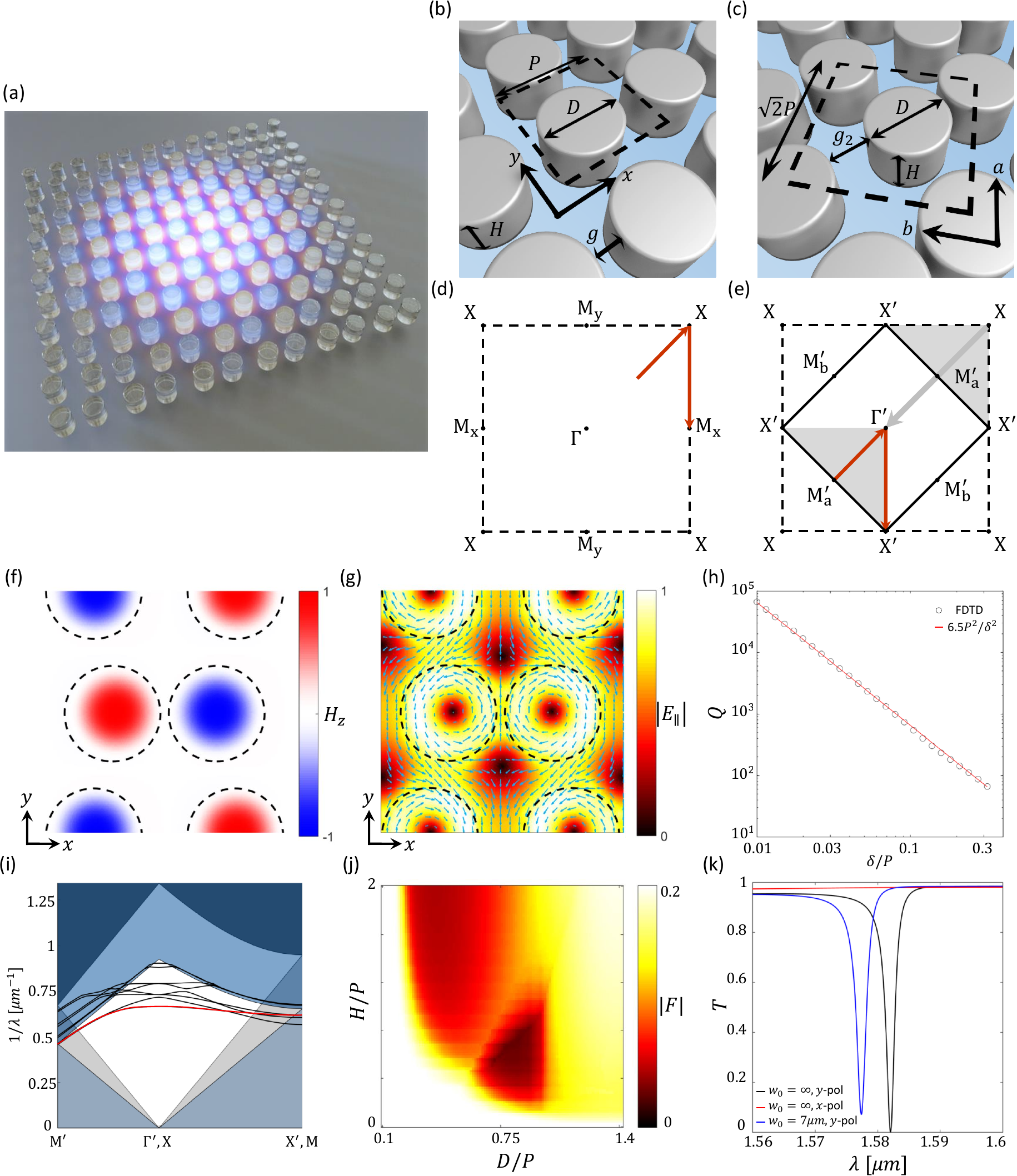}
\caption{\label{fig1}(a) Artistic rendering of a quasi-bound state in the continuum in a periodically perturbed square lattice. (b) Geometry of the unperturbed lattice. (c) Geometry of the perturbed lattice. (d) First Brillouin zone of the unperturbed lattice, with red arrows tracing the path used in the band diagram of (i). (e) First Brillouin zone of the perturbed lattice, showing band folding. (f,g) Out-of-plane magnetic field and in-plane electric field components of the fundamental mode of the perturbed lattice. (h) Dependence of the Q-factor on the perturbation, $\delta = g_2 - g$. (i) Band diagram (with target band highlighted in red) of the folded modes in a finite height PCS with $D = 0.411 \mu m$, $H = 0.295 \mu m$, $P = 0.527 \mu m$. These parameters correspond an operating wavelength of $\lambda = 1.58 \mu m$ with the optimal (minimal) figure of merit $\left \lvert F\right \rvert$ as found by the parameter sweep in (j), in which $|F|$ is mapped for varying $D/P$ and $H/P$. The taller diamond-like window in (i) represents the region of energy-momentum space where the superstrate (air) supports only a single diffractive order ($m=0$); the shorter window represents the same for the substrate (silicon dioxide). (k) Transmittance, $T$, near the fundamental mode frequency of an infinitely periodic device excited by a planewave at normal incidence and of a finite device ($30\mu m\times 30\mu m$) with $\delta = 80 nm$ excited by a Gaussian beam with $e^{-2}$ waist radius of $w_0 = 7 \mu m$. Both devices show $Q \approx 1,000$ and excellent resonance visibility, indicating that the performance of the finite device is maintained despite its small footprint.}
\end{figure*}

We begin by reviewing the design process to create a finite-sized DHCG, we explore a BIC in a diatomic lattice artistically depicted in Fig.~\ref{fig1}(a). Figure~\ref{fig1}(b) and~\ref{fig1}(c) define the geometric parameters of the unperturbed and perturbed lattices, respectively. The two ``atoms'' (here, pillars of Silicon) in the perturbed lattice are identical in height, $H$, and diameter, $D$, and sit in a lattice of period $\sqrt{2}P$, where $P$ is the period of the unperturbed lattice. The perturbation can be quantified as the gaps between atoms: the perturbed gap is $g_2 = g + \delta$, where $g$ is the unperturbed gap ($g = P-D$) and $\delta$ is the perturbation. 
The FBZs of the unperturbed and the perturbed lattices are shown in Fig.~\ref{fig1}(d) and~\ref{fig1}(e), respectively, with high symmetry points defined and the primed coordinates representing the perturbed lattice. The effect of the lattice transformation (taking the period in real space from $P$ to $\sqrt{2}P$ and rotating the basis vectors by $45\degree$) is to shrink the extent of the FBZ and rotate it by $45\degree$. The states belonging to sections of the unperturbed FBZ that lie outside of the new, perturbed FBZ are, by Bloch’s Theorem, equivalent to states within the new FBZ. They are brought into the new FBZ by translation of a reciprocal lattice vector (a process known as Brillouin Zone folding) as depicted graphically in Fig.~\ref{fig1}(e) for the shaded area near the $X$ point. The bound modes that were at the $X$ point are now at the $\Gamma$ point (that is, in the continuum) due to the perturbation.

The new modes brought into the continuum may now couple and produce Fano-like sharp spectral features for normally incident light. By construction, the coupling strength is related to the magnitude of the perturbation. It has been shown~\cite{overvig_dimerized_2018} that the coupling strength for small perturbations is of the order of $\delta$. Since the Q-factor of a sharp resonance is inversely proportional to the square of the coupling strength~\cite{fan_temporal_2003}, a symmetry-protected BIC has a Q-factor governed by~\cite{overvig_dimerized_2018,koshelev_asymmetric_2018}
\begin{equation}
Q=C/\delta^2 \label{Qdelt}
\end{equation}
where the constant $C$ can vary depending on the mode, geometry, materials, and polarization. Figure~\ref{fig1}(f) and~\ref{fig1}(g) show the mode profiles for the fundamental mode depicted in Fig.~\ref{fig1}(a). Figure~\ref{fig1}(h) shows full-wave simulations of the Q-factor of the fundamental mode as a function of perturbation strength, agreeing well with Eq. (\ref{Qdelt}) with $C \approx 6.5 P^2$. Figure~\ref{fig1}(i) contains the band diagram for the perturbed structure calculated by the planewave expansion method (PWEM) using the supercell method, with high symmetry points defined relative to both the unperturbed and perturbed lattices. The modes are calculated in the unperturbed structure following the red arrows in Fig.~\ref{fig1}(d), and then artificially folded into the FBZ.

The band structure of the perturbed PCS in Fig.~\ref{fig1}(i) can help predict the accuracy of Eq. (\ref{Qdelt}) for finite devices. In an infinite device, a planewave corresponds to a single state (for instance, a mode at the $\Gamma$ point) and the band curvature is irrelevant. However, a finite device excited by a Gaussian beam will behave as some combination of responses excited by the planewaves composing that Gaussian beam. One simple model for predicting the behavior of a finite device is to perform a weighted sum of the spectra corresponding to the constitutive planewaves~\cite{overvig_dimerized_2018}. We model a band by a Taylor expansion about the $\Gamma$ point, $\omega_{res}(k) = \omega_{0} + bk^2$, where $\omega_0$ is the angular frequency of the mode at $k = 0$ and $b = \frac{1}{2}\left.\frac{\partial^2\omega_{res}}{\partial k^2} \right|_{k=0} $. A Gaussian beam with a characteristic spread in wavevector of $\Delta k$ will excite a characteristic spread of frequencies $\Delta\omega = b\Delta k^2$. It is natural to expect that if this spread of frequencies is larger than the linewidth of the resonance, $d\omega$, excited in an infinite device by a planewave, the spectral feature will be washed out, lowering the observed $Q$ and invalidating Eq. (\ref{Qdelt}). This suggests a constraint 
\begin{equation}
Q = \frac{\omega_0}{d\omega} \leq \frac{\omega_0}{\Delta \omega} = \frac{\omega_0}{b\Delta k^2}. \label{Qbk}
\end{equation}
In other words, there is an upper limit on the Q-factor attainable in a finite device due to the band curvature near the $\Gamma$ point.

While this simple model does not account for all of the possible finite size effects (e.g., edge effects and a more complex modal structure), the derived constraint suggests that optimizing the band flatness will tend to allow for the most compact devices. In particular, the factor to minimize is $F = \left \lvert b\right \rvert/\omega_0$, which serves as a figure of merit when designing a device by computing its band structure. Figure~\ref{fig1}(j) maps $F$ as calculated for a variety of diameters and heights (relative to the period) of Silicon pillars sitting on a Silicon Dioxide substrate. While the curvature is different along $\Gamma - M'$, or the $k_x$ direction, compared to along $\Gamma - X'$, or the $k_a$ direction, this band is limited by its curvature in the $k_x$ direction; we therefore restrict the calculation of $F$ to the band along the $k_x$ direction. 

We choose a design with the smallest $F$ according to Fig.~\ref{fig1}(j) and scale its geometrical parameters by a factor $\lambda/\lambda_{res}$ such that the operating resonant wavelength is $\lambda = 1.58\mu m$ for a calculated resonant wavelength, $\lambda_{res}$. Figure~\ref{fig1}(k) shows transmittance spectra calculated by full-wave simulations of an infinitely periodic device excited by a planewave of either $x$ or $y$ polarization, demonstrating that this coupling only occurs for $y$ polarization. Figure~\ref{fig1}(k) also shows a transmittance spectrum of a device of finite size ($30 \mu m \times 30 \mu m$) excited by a Gaussian beam with a waist radius of $w_0 = 7 \mu m$ calculated by full-wave simulations. The spectral feature remains intact, confirming that the flat band in Fig.~\ref{fig1}(i) determined through the optimization shown in Fig.~\ref{fig1}(j) allows for compact devices with moderately high $Q \approx 10^3$.

Figure~\ref{fig1} overviews the design process of a compact optical device (a two-dimensional DHCG) supporting a sharp spectral feature due to a quasi-BIC. However, this process represented just one high symmetry mode, and explored the behavior as a result of only one specific perturbation. This behavior was shown to be weakly dependent on incident angle (Fig.~\ref{fig1}(i)), but strongly dependent on incident polarization (Fig.~\ref{fig1}(k)). The key result of this paper is a theoretical description allowing prediction of the polarization behavior (selection rules) of all high symmetry modes and perturbations. With this result, which we call the ``catalog of selection rules", the design process overviewed in Fig.~\ref{fig1} can be summarized as having three steps: (1) Choose a high symmetry mode for its real-space properties (e.g., for its field overlap with the high index material); (2) Optimize the band structure by tuning the parameters in unperturbed structure (as in Fig.~\ref{fig1}(i)); (3) Choose a proper perturbation according to the desired selection rules (e.g., targeting $y$ polarization). The catalog serves as a comprehensive guide for step (3), clarifying the wealth of options in conjunction with the choice in step (1) of the desired high symmetry mode; it thereby provides a high-level roadmap for this three step design process resulting in a PCS that confines light in both space and time. This design scheme may be further coupled with computational inverse design techniques~\cite{molesky_inverse_2018} to reduce the dimensions of the design parameter space to be explored.

\subsection{Classification of High Symmetry Modes}
\label{B2}
\begin{figure*}
\includegraphics[width=1.7\columnwidth]{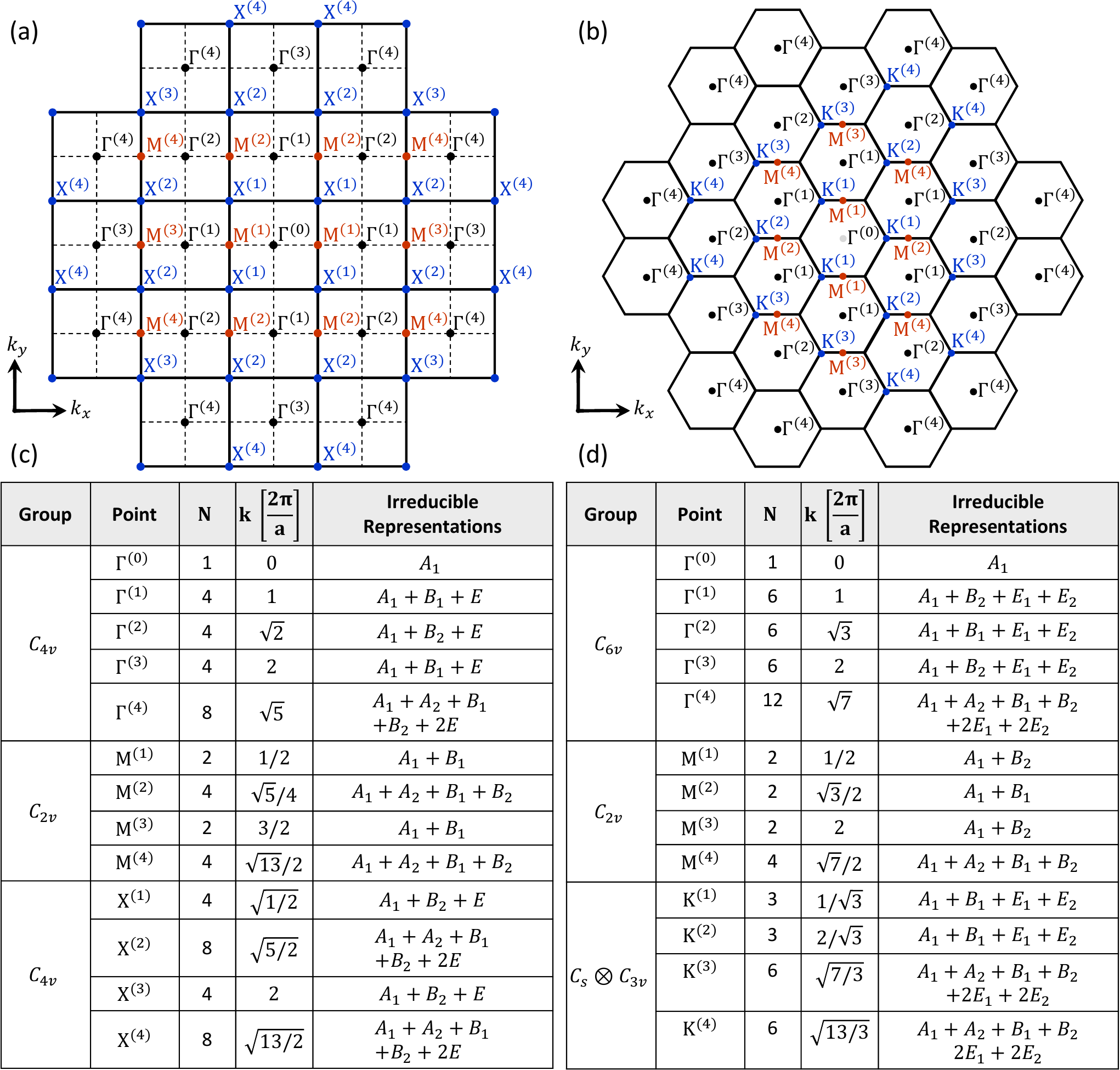}
\caption{\label{fig2}Extended zone scheme mode classification. (a,b) Extended zones in reciprocal space of the square and hexagonal lattices. (c,d) Mode classification tables for the square and hexagonal lattices detailing the point group (column labeled ``Group''), extended zone (column labeled ``Point''), number of modes (column labeled ``$N$''), characteristic wavevector of the planewave (column labeled ``$k$''), and the irreducible representations (mode symmetries) present at each extended zone for each of the six high symmetry mode types.}
\end{figure*}
The first step to determining the selection rules of perturbed PCSs is to classify the modes present. Since the selection rules arise from symmetry breaking, a mode classification scheme employing the symmetries of the allowed modes is the natural choice. Although the final devices of interest are three-dimensional in nature (having a finite thickness in the out-of-plane direction, $z$), it considerably simplifies the analysis to begin with Maxwell’s equations in two dimensions. In this case, Maxwell’s curl equations decouple into two separate sets of three equations, each set defining modes characterized by either the out-of-plane magnetic field, $H_z$, (referred to as TE modes) or the out-of-plane electric field, $E_z$ (referred to as TM modes). Each mode is then definable by this single field component. We therefore select, review, and carry out a Group Theory approach detailed in Ref.~\cite{sakoda_optical_2005} to classify the modes by in-plane symmetries of the out-of-plane field component. We note that this Group Theory analysis is valid for any materials system, for instance, an array of Silicon pillars, holes in a Silicon slab, or even a metallic structure. For convenience, and comparison to conventional metasurfaces, we first consider arrays of Silicon pillars. But the resulting selection rules are immediately transferable to any other materials system.

\begin{figure*}
\includegraphics[width=2.1\columnwidth]{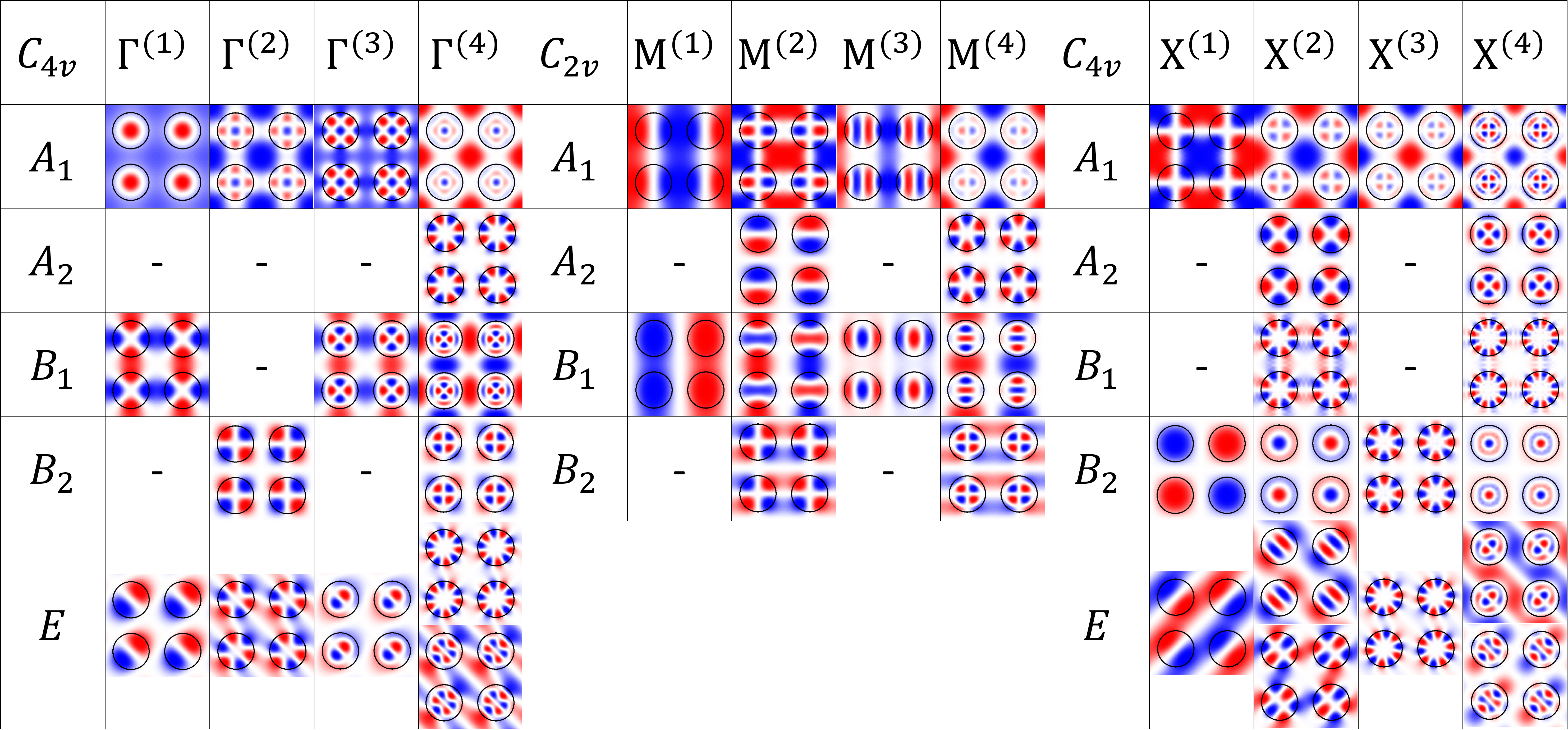}
\caption{\label{fig3}Modes at the high symmetry points in the square lattice, classified by in-plane symmetries (column-wise) and extended zone (row-wise). The three tables correspond to the modes at the $\Gamma$, $M$, and $X$ points, respectively. Modes are calculated by planewave expansion method for the electric field out of plane (TM modes); an analogous set exists with magnetic field out of plane (TE modes).}
\end{figure*}

\begin{figure*}
\includegraphics[width=2.1\columnwidth]{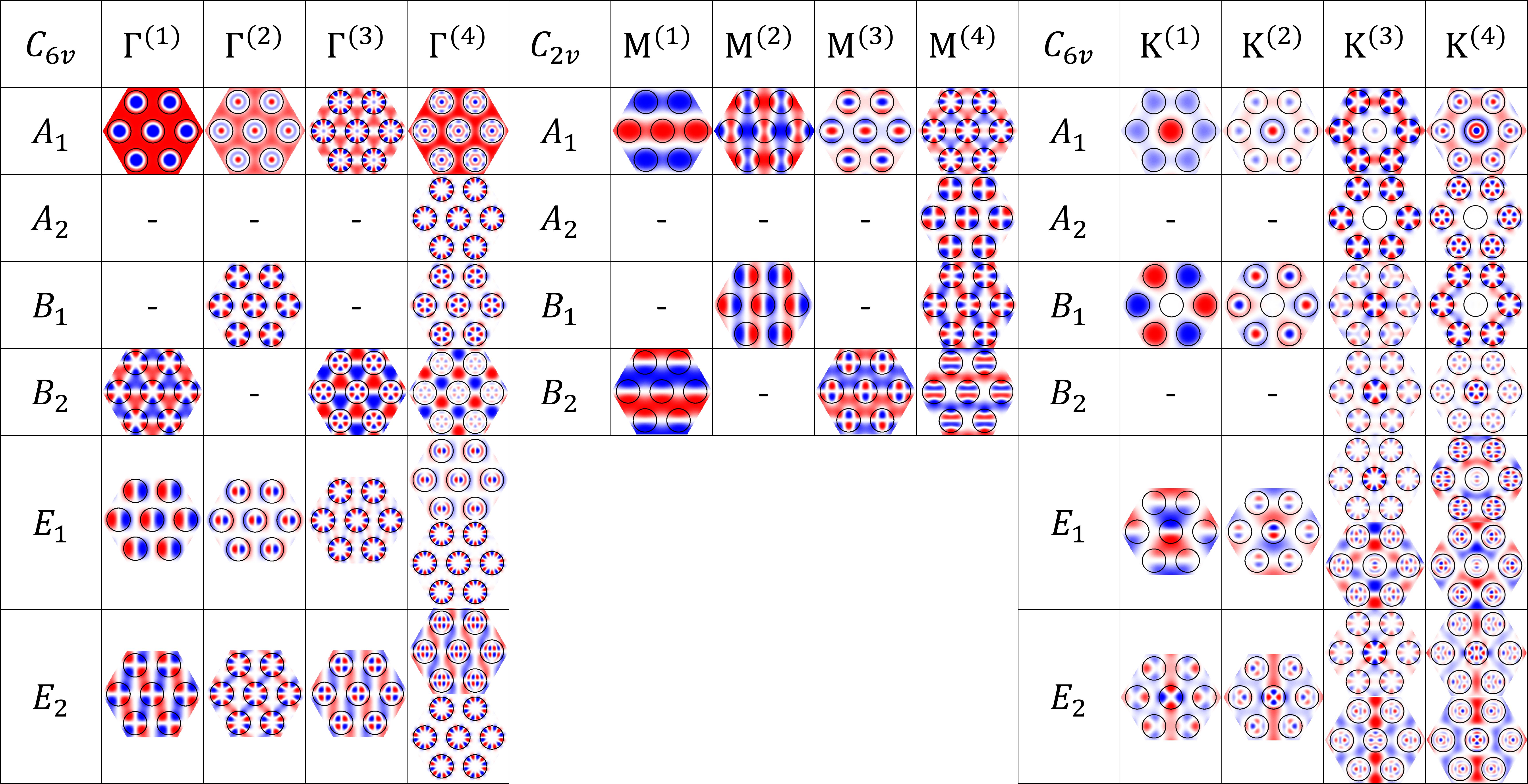}
\caption{\label{fig4}Modes at the high symmetry points in the hexagonal lattice, classified by in-plane symmetries (column-wise) and extended zone (row-wise). The three tables correspond to the modes at the $\Gamma$, $M$, and $K$ points, respectively. Modes are calculated by planewave expansion method for the electric field out of plane (TM modes); an analogous set exists with magnetic field out of plane (TE modes).}
\end{figure*}

Since the fields exist in a periodic lattice, they are characterized by planewaves with magnitudes and directions corresponding to high symmetry points of the reciprocal lattice. When the index contrast is low, this characterization is excellent; when the index contrast is large (such as a Silicon and air system), significant deviations in resonant frequencies occur relative to the low index contrast systems, but the symmetries of the possible modes remain unchanged. The modes can therefore be studied with reference to the extended zone scheme.

Figure~\ref{fig2}(a) depicts the extended zone scheme for a square lattice, with notable high symmetry points marked. In particular, the $\Gamma$ points, $M$ points, and $X$ points are labeled by an index pertaining to their distance from the origin, $\Gamma^{(0)}$. These points have point group symmetries $C_{4v}$, $C_{2v}$, and $C_{4v}$, respectively (see Appendix \ref{A} for the character tables and other relevant Group Theory tables), and the modes decomposable by planewaves corresponding to these points are describable by these point groups. These three sets of points are the highest symmetry points in the reciprocal lattice, and therefore correspond to the modes of interest in the square lattice. The three analogous sets of points in the hexagonal lattice (Fig.~\ref{fig2}(b)) are the $\Gamma$ points, $M$ points, and $K$ points.

A Group Theory approach~\cite{sakoda_symmetry_1995,sakoda_optical_2005,yu_all-dielectric_2018} predicts the number and nature of the modes from each set of high symmetry points in the extended zone scheme. Figure~\ref{fig2}(c) contains a table summarizing the modes possible at each of the high symmetry points in the square lattice. The degeneracy of a set of high symmetry points, $N$, is also the number of modes corresponding to that set. The magnitude of the wavevectors, $k$, of the planewaves of a set will correspond to the expected eigenfrequencies of the modes (however, as noted above, this correspondence is poor in high-index contrast systems). Lastly, the irreducible representations describe the mode symmetries. That is, modes that ``transform like" (share all the symmetries of) each irreducible representation listed in an extended zone will be present at that extended zone. Note that the $E$ irreducible representations are doubly degenerate, and so account for two modes.

Figures~\ref{fig3} and~\ref{fig4} depict the TM modes from the first four extended zones of each high symmetry point in the square and hexagonal lattices, respectively. An analogous set exists for TE modes, identical in symmetries (in $H_z$ instead of $E_z$) but spatially distorted and differing in eigenfrequency. The modes are organized by the extended zone order (columns) and irreducible representation (rows). Reference to the relevant character tables (Appendix \ref{A}, Fig.~\ref{Group1}(a)) shows that modes labeled by a given irreducible representation transform the same way as the corresponding row in the character table: a $1$ in a column of this row means the mode will be symmetric under the class of operations of that column; a $-1$ means anti-symmetric; a $0$ means not symmetric; and a magnitude of $2$ signifies that the mode is degenerate. 

Finally, the out-of-plane property of the modes is characterized by the order, $n$, or number of anti-nodes, per atom of the PCS in the $z$ direction. The inclusion of out-of-plane characteristics captures all the relevant features of the modes within the scope of this paper if the PCS has mirror symmetry about an $xy$ plane. However, two-dimensional PCSs with a substrate are known to exhibit chiral behavior: incident circularly polarized light can behave in a manner depending on the handedness~\cite{wu_spectrally_2014,eismann_exciting_2018}. The chiral effects of a substrate and vertical symmetry breaking are beyond the scope of this paper, and represent a fruitful avenue for future research. We restrict ourselves to PCSs composed of vertically extruded 2D lattices, and we find that the presence of a low-index substrate (such as glass) generally has little practical effect of this kind (and so can be ignored).

With the in-plane and out-of-plane features of each mode classified, we are motivated to provide a naming scheme. We call a mode:
\begin{equation}
\psi_{L,S}^{m,n}, \label{mn}
\end{equation}
where $\psi$ is TM or TE if the mode is characterized by $E_z$ or $H_z$, respectively, $L$ signifies the reciprocal lattice point (e.g., $\Gamma$), $S$ is the irreducible representation (e.g., $A_1$), $m$ is the extended zone order, and $n$ is the out-of-plane order. For instance, the mode in the $B_2$ row and $X^{(1)}$ column in Fig.~\ref{fig3}, with a single out-of-plane anti-node per unit cell of the PCS would be called $\text{TM}_{X,B_2}^{1,1}$, which is the lowest frequency $E_z$ mode in this square lattice. $\text{TE}_{X,B_2}^{1,1}$ is the mode explored in Fig.~\ref{fig1}.

Importantly, we discuss the relationship of the 2D description of the modes and the modes of a finite-thickness PCS, which we assume throughout this Article is simply extruded (its cross-section is invariant) in the $z$ direction. The modes depicted in Fig.~\ref{fig3} and Fig.~\ref{fig4} may be considered as the modes traveling in the $z$ direction in a semi-infinite 2D photonic crystal. In this case, these eigenmodes are described by folding the eigenmodes of an unpatterned isotropic medium (i.e., planewaves), while in a finite PCS, as discussed in \cite{sakoda_optical_2005}, the eigenmodes are described by folding the eigenmodes of the unpatterned slab. While the underlying bases are distinct, the resulting modal symmetries are identical. However, polarization mixing occurs in the PCS as a result of the finite thickness, yielding modes that are quasi-TE and quasi-TM, rather than pure. But due to the vertical extrusion, and for normally incident light, the new polarization components cannot introduce or destroy any symmetries, which are described by the point group; that is, the TM (TE) components of the quasi-TE (quasi-TM) modes contain equivalent symmetry properties to the analogous pure-TE (pure-TM) modes, and therefore have no bearing on the selection rules for normally incident light. We refer to quasi-TE (quasi-TM) modes as simply TE (TM) for this reason, and are free to treat the PCS modes as equivalent to their 2D counterparts for the purposes of studying the selection rules at normal incidence. 

The modes of a PCS also may have additional, out-of-plane symmetries compared to their 2D counterparts. If there is a $xy$ mirror-plane (at the center of the PCS), then the point groups used here are not sufficient to describe the modes. For instance, the modes of the hexagonal lattice are described by the dihedral group $D_{6h} = C_{6v}\otimes C_{1h}$, where the $C_{1h}$ group accounts for whether the modes are symmetry or anti-symmetric with respect to that $xy$ mirror plane. This symmetry determines, in part, the ``handedness'' of the Fano resonance~\cite{fan_temporal_2003} (that is, whether the reflection peak occurs at a redder or bluer wavelength than the reflection dip). However, it has no bearing on the selection rules at normal incidence, and may therefore be ignored for our purposes. Furthermore, in the most practical scenario a low-index substrate is present, breaking this mirror symmetry and leaving only in-plane symmetries. Since inclusion of the mirror plane needlessly doubles the number of modal labels, we exclude it. 

Instead, we use the modal index $n$ to refer the out-of-plane characteristics of the PCS modes. When $n$ is odd (even), the decay symmetry is even (odd), meaning that the mode naming scheme in Eq.~\ref{mn} contains the relevant information about the handedness of the asymmetric lineshape (note that the decay symmetries hold approximately even in the presence of a low-index substrate). Consequently, there is a \textit{1:many} correspondence between the modes of a semi-infinite photonic crystal and a PCS: each mode in Figs.~\ref{fig3} and \ref{fig4} has identical in-plane symmetries to \textit{many} modes in the PCS that differ only out-of-plane according to the modal index $n$. Such a relationship between accidental BICs supported by an HCG and the vertically propagating waveguide modes of the corresponding 1D waveguide array is described in \cite{chang-hasnain_high-contrast_2012}, wherein a round-trip phase condition of the vertically propagating modes predicts the dispersion of the BICs; here, the integer multiple of $2\pi$ picked up upon a round trip is the modal index $n$.

Finally, we note that the $K$ point modes in the hexagonal lattice are more complex than the other five high symmetry modes. In particular, the $K$ point in the unperturbed lattice has point symmetry of $C_{3v}$, as evident in Fig.~\ref{fig2}(b). However, there are two identical sets of these $K$ points; the set not pictured in Fig.~\ref{fig2}(b) can be obtained by reflection about the $k_y$ axis. As demonstrated in Fig.~\ref{fig1}, the modes only become BICs once folded to the $\Gamma$ point by an appropriate perturbation (depicted in Fig.~\ref{fig5}). For $K$ point modes, the analogous perturbation results in a triatomic lattice and therefore triples the number of modes at the new $\Gamma$ point compared to the unperturbed monatomic lattice. At the perturbed $\Gamma$ point, a set of modes originating from each set of $K$ points will mix in pairs. The symmetries of the mixed modes are described by the direct product group $C_s \otimes C_{3v}$, corresponding to the relationship of the two sets of $K$ points. This direct product group is isomorphic to (shares the same character table as) the group $C_{6v}$, allowing the modes to be named in $C_{6v}$. Naming the modes according to $C_{6v}$ is inconsistent with the modes in an unperturbed lattice at a single $K$ point, but consistent with the modes upon folding to the $\Gamma$ point and mixing in the relevant perturbed lattice. Since our goal is to study these modes in the perturbed lattice, defining the modes in $C_{6v}$ is the more fruitful choice.

\section{Selection Rules}
\label{SelectionRules}

By proper periodic perturbation, any of the six classes of high symmetry modes can be accessed from free space if additional symmetry constraints are satisfied. These symmetry constraints can be treated with a Group Theory approach, and result in a catalog detailing how each high symmetry mode classified above couples to free space under a given planar perturbation. In the following, we identify six lattice types chosen to target the six high symmetry modes (Fig.~\ref{fig5}), list all the degenerated space groups compatible with those lattice types (Fig.~\ref{fig6}), and then derive the selection rules for every case (exemplified in Figs.~\ref{fig7} and~\ref{fig8}). The resulting catalogs (Figs.~\ref{fig9} and~\ref{fig10}) can be used as a high-level roadmap in the design of planar photonic devices. 

\subsection{Target Space Groups}
\label{S1}
\begin{figure*}
\includegraphics[width=1.6\columnwidth]{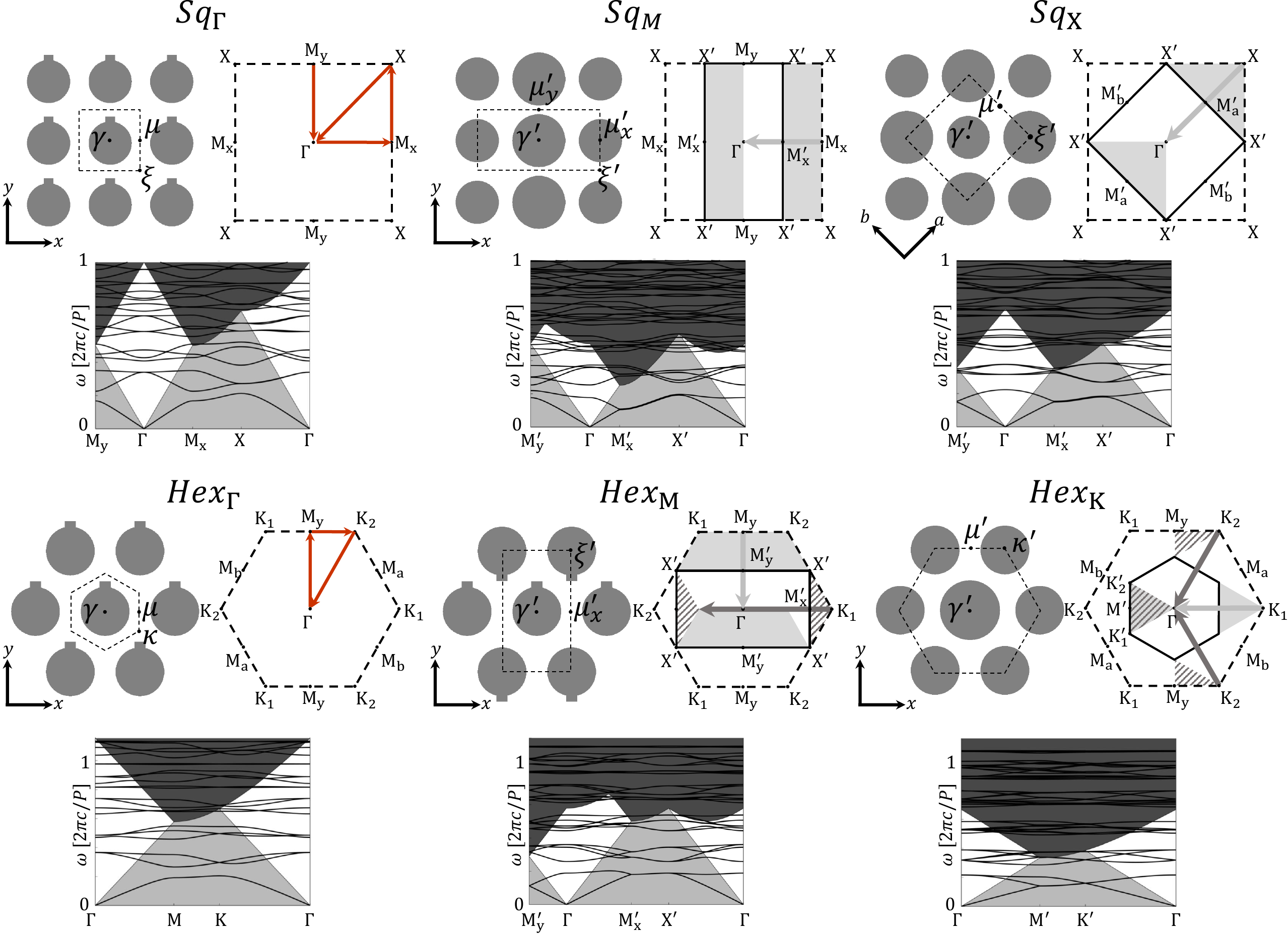}
\caption{\label{fig5}Six lattices target six distinct high symmetry modes. They are named for their lattice family ($Sq$ for square lattices and $Hex$ for hexagonal lattices) and the high symmetry mode they uniquely target (e.g., $Sq_M$ folds the $M$ point modes of a square lattice into the continuum by a period doubling perturbation). An example unit cell with a perturbation is given with high symmetry points defined ($\gamma, \mu, \xi$, and $\kappa$). The FBZ is also given with high symmetry points defined ($\Gamma, M, X$, and $K$), dashed lines denoting the FBZ of the unperturbed lattice and solid lines that of perturbed lattice. Lastly, an example band diagram is shown for infinitely tall PCSs for the TM polarization case, showing generally the presence of flat bands at the $\Gamma$ point and band folding in the relevant cases. The red arrows in the $Sq_\Gamma$ and $Hex_\Gamma$ FBZs depict a representative path taken through the FBZ for the band diagrams. Modes in the light shaded area are bound. Modes in the white areas are in the continuum accessible to a single diffractive order (``0\textsuperscript{th} order diffraction'') and are the focus of this paper. Modes in the dark shaded area are higher order diffractive modes.}
\end{figure*}

The six types of high symmetry modes described above motivate six types of lattices, each one uniquely targeting one of the six high symmetry mode types. For each of these lattices, an exhaustive list of lattices with lower symmetry attainable by planar perturbation is explored. The symmetry degeneration from higher symmetry to lower symmetry will constrain which polarization, if any, may couple to free space for each high symmetry mode.

The six lattice types, depicted in Fig.~\ref{fig5}, are named based on the modes they target and whether they begin with square or hexagonal symmetries. For instance, the $Sq_\Gamma$ is a monatomic photonic crystal with a square lattice where the perturbation has periodicity equal to that of the unperturbed lattice. This lattice is labeled by $\Gamma$ because it supports none of the other types of modes of interest supported by the square lattice (that is, $M$ and $X$ modes) in the continuum. Figure~\ref{fig5} (top left) depicts an example real space lattice, First Brillouin Zone (FBZ), and band diagram for the $Sq_\Gamma$ lattice. The white region in the band diagram is the region of the continuum of interest, wherein only the 0\textsuperscript{th} diffractive order is allowed. We constrain ourselves to the area near the $\Gamma$ point of the white region, where the symmetry-protected BICs can produce sharp spectral features described above.

The $Sq_M$ lattice (top middle of Fig.~\ref{fig5}), on the other hand, is a photonic crystal with perturbations with periodicity double that of the unperturbed lattice in a single direction. This period doubling (in the $x$ direction in Fig.~\ref{fig5}) halves the extent of the FBZ in the $k_x$ direction. The shaded portion outside the new FBZ is then translated into the FBZ by a reciprocal lattice vector. As a result, the $M$ point of the unperturbed lattice overlaps with the $\Gamma$ point, bringing the $M$ point modes into the continuum in an analogous way described in the example in Fig.~\ref{fig1}. This Brillouin Zone folding also changes the shape of the 0\textsuperscript{th} order diffraction region of the band diagram. The $\Gamma$ point will now have both the modes at the unperturbed $\Gamma$ point as well as at the unperturbed $M$ points. The $Sq_M$ lattice is the only lattice in Fig.~\ref{fig5} to bring the $M$ point modes of a square lattice into the continuum, motivating its name. The remaining lattices target $X$ modes of the square lattice ($Sq_X$, which is the lattice type explored in Fig.~\ref{fig1}), and $\Gamma$, $M$, and $K$ point modes of the hexagonal lattice ($Hex_\Gamma$, $Hex_M$, and $Hex_K$, respectively) in an analogous way. Notably, two distinct regions are folded into the FBZ of the $Hex_K$ lattice: as discussed above, two sets of modes are folded to the $\Gamma$ point, one from each distinct $K$ point.

We note that the six lattices chosen in Fig~\ref{fig5} are not an exhaustive set: lattices with any number of atoms per unit cell are possible. Ordering this list of lattices by number of atoms per unit cell, the six chosen lattices are the lowest order lattices uniquely targeting the six high symmetry modes of interest. Appendix~\ref{B} describes three examples of higher order lattices. The approach described in what follows may be used to generate the catalog for any higher order lattice.

\begin{figure}
\includegraphics[width=1\columnwidth]{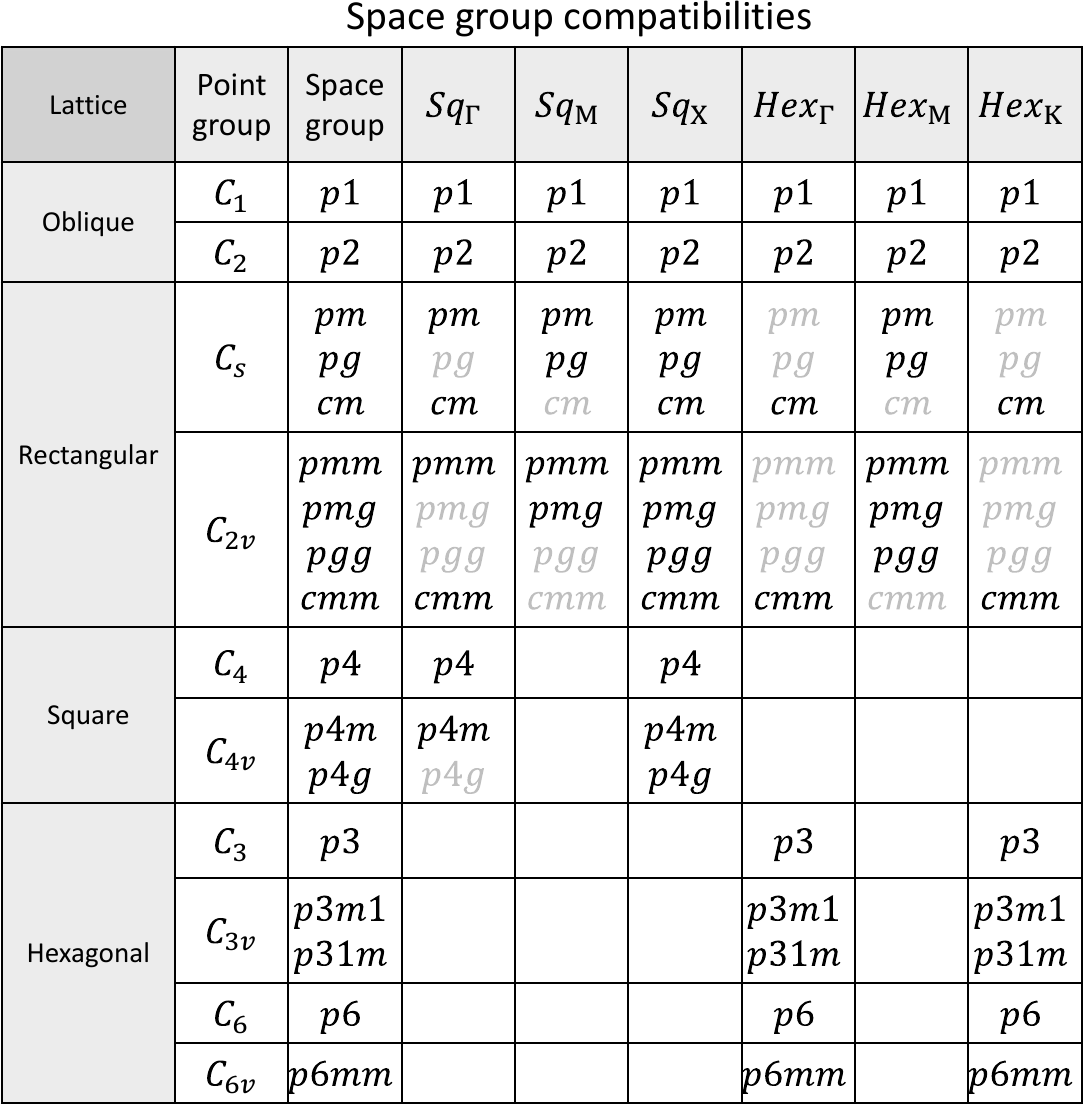}
\caption{\label{fig6}Space group compatibility table. Different lattice families (column 1) are compatible with various point groups (column 2), each of which can be further subdivided into the $17$ ``wallpaper groups'' (or two-dimensional space groups, column 3). The remaining columns track the compatible space groups of each degenerated lattice studied. A blank entry means that corresponding space group is excluded due to a mismatch in lattice family; a greyed entry means it is excluded because it has an incompatible glide symmetry.}
\end{figure}

Next, the space groups of degenerated lattices that are compatible (attainable through perturbation) with each lattice are determined and reported in Fig.~\ref{fig6}, the space group compatibility table. First, all $17$ ``wallpaper groups'' are listed and categorized by the compatible lattice family (e.g., ``Rectangular''). The point group of each of these is given for reference. Then, for each target lattice type (e.g., $Sq_M$) the space groups compatible with the lattice class (i.e., square or hexagonal) are listed. For instance, the space group $p6mm$ requires hexagonal tiling, and is therefore omitted as a possibility for any $Sq$ lattice. Likewise, a $Hex$ lattice cannot be perturbed into a square lattice without distortion of the lattice vectors, but it can be perturbed into a rectangular lattice. The space groups of the square lattice family are therefore omitted from all $Hex$ lattices, but those of the rectangular lattice family are not. 

Next, the glide reflection operation (a reflection and a translation by a fraction of a unit cell) is tested for each type of lattice. Glide reflections are present in only some two-dimensional space groups and are not compatible with all of the six target lattice types. As an example, it is quickly found by inspection that the $Sq_\Gamma$ lattice does not support glide symmetries in directions other than along the diagonals (more rigorously, in the language of crystallography, monatomic PCSs are incompatible with non-symmorphic space groups). This excludes the space groups $pg$, $pgg$, and $pmg$, which are correspondingly greyed out in the column for $Sq_\Gamma$. For the same reason, for the $Sq_M$ lattice, glide planes along the direction where the lattice is unperturbed (and therefore monatomic) are incompatible (the $y$ direction in Fig.~\ref{fig5}). Additionally, all diagonal glides are incompatible for the $Sq_M$ lattice because they correspond to reflection axes that are not included in the point group of any $Sq_M$ lattice. This excludes $cm$, $pg$, and $cmm$ for $Sq_M$, which are greyed out accordingly. There are no such constraints for the $Sq_X$ lattice, which can be degenerated into a lattice of any space group (except the hexagonal ones). Similar arguments can be made for the $Hex$ lattices, and the results are reported in Fig.~\ref{fig6}.

Finally, it must be noted that there exist multiple high symmetry points in each real space lattice. These are given names in Fig.~\ref{fig5} for each case. For instance, the $Sq_\Gamma$ lattice has two points having the full symmetry of the $C_{4v}$ point group, named $\gamma$ and $\xi$. Both are perfectly acceptable to choose as the reference point: in the mode naming scheme in Sec.~\ref{B2}, the $\xi$ point is the reference point, but the modes may all be renamed according to the $\gamma$ point if desired. Similarly, a degenerated space group may choose either of these points to have in common with the unperturbed lattice. Generally speaking, every degenerated space group may be tried with each of the high symmetry points in common with the unperturbed lattice, thereby allowing for more than a single unique example of each space group in each lattice. For instance, there are three distinct $cmm$ space groups in the $Sq_X$ catalog (see Fig.~\ref{fig9}): one with the $\gamma$ point as the high symmetry point in common, one with $\xi$, and the last with $\mu$ (which is the space group of the perturbation in Fig.~\ref{fig1}). As shown in Fig.~\ref{fig9}, though these have identical space groups, they do not have identical selection rules because they are attained through distinct perturbations. Therefore, to determine all of the unique symmetry degenerations possible, an attempt is made to construct each compatible space group (Fig.~\ref{fig6}) with each high symmetry point in common between the unperturbed and perturbed lattice. The successful attempts comprise the set of all degenerated lattices compatible with those chosen in Fig.~\ref{fig5}. This proof by exhaustion is omitted here. For each of these degenerated lattices, the modes derived in the previous section can be studied, and their selection rules derived. The results are tabulated in Figs.~\ref{fig9} and~\ref{fig10} following the methods derived in Secs.~\ref{S2} and~\ref{S3}.

\subsection{Deriving the Coupling Condition}
\label{S2}

To derive the selection rules reported in Figs.~\ref{fig9} and~\ref{fig10}, we study the end-fire coupling of free-space light normally incident to a semi-infinite 2D photonic crystal supporting the modes depicted in Fig.~\ref{fig3} and Fig.~\ref{fig4}. As described in Sec.~\ref{B2}, there is a \textit{1:many} correspondence preserving in-plane symmetries between the modes excited by this end-fire coupling and the modes excited in a finite-height PCS (under the assumptions that the cross-section is invariant along the $z$ direction, whose substrate and superstrate are isotropic media). At normal incidence, the two cases therefore have identical selection rules with respect to in-plane symmetry breaking, and so we may proceed with the simpler case of end-fire coupling. The multiplicity of modes of the PCS differentiated by the modal index $n$ have identical selection rules; it is only the in-plane symmetries that are relevant. In particular, we determine under which conditions the end-fire coupling coefficient, $\gamma_e$, is non-vanishing:
\begin{equation}
\gamma_e \propto \iint_A [\boldsymbol{E}_{inc}^*\times \boldsymbol{H}_{mode} + \boldsymbol{E}_{mode}\times \boldsymbol{H}_{inc}^*]\cdot \hat{\boldsymbol{z}}   \,dx\,dy, \label{g1}
\end{equation}
where $\hat{\boldsymbol{z}}$ is the unit vector in the $z$ direction and bold symbols refer to vector quantities. We exclude a normalization factor for simplicity (it does not affect whether $\gamma_e$ vanishes) and evaluate the integral over the area, $A$, of a unit cell.
Here, we take the incident field to be a normally incident planewave, with electric field
\begin{equation}
\boldsymbol{E}_{inc}^* = 
\begin{bmatrix}
A_x \\
A_y \\
0
\end{bmatrix},
\end{equation}
and magnetic field
\begin{equation}
\boldsymbol{H}_{inc}^* = \frac{1}{\eta_0}
\begin{bmatrix}
A_y \\
A_x \\
0
\end{bmatrix},
\end{equation}
where $\eta_0$ is the impedance of free space.
And the mode has electric field 
\begin{equation}
\boldsymbol{E}_{mode} = e^{i(\beta z-\omega t)} 
\begin{bmatrix}
E_x \\
E_y \\
E_z
\end{bmatrix},
\end{equation}
and magnetic field 
\begin{equation}
\boldsymbol{H}_{mode} = e^{i(\beta z-\omega t)}
\begin{bmatrix}
H_x \\
H_y \\
H_z
\end{bmatrix}, 
\end{equation}
where $\beta$ is the propagation constant satisfying the dispersion relation $\omega = c\beta$. Evaluating the cross-products in Eq.~(\ref{g1}), the free-space coupling coefficient is written as
\begin{equation}
\gamma_e \propto \iint_A [A_x(\eta_0H_y + E_x) + A_y(\eta_0H_x + E_y)] \,dx\,dy. \label{g2}
\end{equation}
Using Maxwell’s curl equations, the in-plane components ($E_x, E_y, H_x, H_y$) are replaced with the out-of-plane components ($E_z, H_z$) to both simplify the equation and allow the previous mode classification scheme (based on the out-of-plane field components) to straightforwardly apply. The resulting form is 
\begin{eqnarray}
\gamma_e \propto \iint_A \big[ A_x(c_1\partial_xE_z+c_2\partial_yH_z) + \nonumber \\*
A_y(c_1\partial_yE_z+c_2\partial_xH_z)\big] \,dx\,dy \label{g3}
\end{eqnarray}
where
\begin{equation}
c_1 = \frac{1}{i\beta}\frac{1+\varepsilon_r(x,y)}{1-\varepsilon_r(x,y)} \label{c1}
\end{equation}
and
\begin{equation}
c_2 = \frac{2\eta_0}{i\beta}\frac{1}{1-\varepsilon_r(x,y)} \label{c2}
\end{equation}
with the replacement $\varepsilon_r(x,y) = \varepsilon(x,y)/\varepsilon_0$ as the relative permittivity. This can be written more compactly as
\begin{equation}
\gamma_e \propto 
\left \langle 
\begin{bmatrix}
A_x & A_y
\end{bmatrix} \cdot
\begin{bmatrix}
c_1\partial_x & c_2\partial_y \\
c_1\partial_y & c_2\partial_x
\end{bmatrix} \cdot
\begin{bmatrix}
E_z \\
H_z
\end{bmatrix}
\right \rangle, \label{g4}
\end{equation}
where angled brackets indicate integration over a unit cell. While it is possible to proceed with this form by considering the symmetries of each component, it is considerably simpler and more informative to reduce this to individual choices of incident polarization (e.g., choose $A_x=0$) and mode type (i.e., choose either TM modes or TE modes). In this case, we write
\begin{equation}
\gamma_e \propto \iint_A c_j\partial_i\psi \,dx\,dy \label{g5}
\end{equation}
where $\psi$ is TM or TE, $\partial_i$ refers to the partial derivative in a relevant high-symmetry direction ($i=x,y,a,b$), and $c_j$ is $c_1$ when $\psi$ is TM and $c_2$ when $\psi$ is TE. 

\subsection{Determining the Selection Rules}
\label{S3}

We now apply Eq. (\ref{g5}) to the modes supported by a 2D photonic crystal. The modes in the unperturbed lattice can be described as the eigenvectors of an eigenvalue equation
\begin{equation}
H^0\psi^0 = \mathcal{E}^0\psi^0 \label{eig0}
\end{equation}
where the superscript marks reference to the unperturbed eigenvalue problem. We are interested in particular in the $\psi^0$ that are uncoupled to free space (i.e., $\psi^0$ for which the integral in Eq. (\ref{g5}) vanishes) due to symmetry. To proceed we apply perturbation theory to determine any non-vanishing terms present in the generalized eigenvalue problem of a degenerated lattice:
\begin{equation}
H\psi = \mathcal{E}\psi \label{eig}
\end{equation}
where $H = H^0 + V$ is perturbed by the perturbation operator, $V$, and $\psi = \psi^0 + \psi^1$ is the perturbed field profile with $\psi^1$ as the first order correction. First order perturbation theory gives the form of the $n$\textsuperscript{th} mode $\psi_n^1$ as
\begin{equation}
\psi_n^1 = \sum_{m\neq n} \frac{\left \langle(\psi_m^0)^*V\psi_n^0\right \rangle}{\mathcal{E}_m-\mathcal{E}_n}\psi_m^0. \label{psi1}
\end{equation}
That is, the perturbed portion of the field is a superposition of the unperturbed fields. (Note that Eq. (\ref{psi1}) is the non-degenerate form of perturbation theory, but it can be applied to degenerate states as well if the ``correct'' orthogonal linear combination of states is known ahead of time. Since these will correspond to a high symmetry direction of the perturbed lattice, they are easy to predict; we therefore use this form to apply to degenerate modes.) However, inspection of the coefficient, $\left \langle(\psi_m^0)^*V\psi_n^0\right \rangle$, in front of each $\psi_m^0$ reveals that not all $\psi_m^0$ will contribute: many will vanish due to symmetry.

The process of determining $\psi_m^0$ that contribute can be clarified and expedited in the language of Group Theory. Specifically, if we can determine the irreducible representations of each factor within the integrand, we can find the symmetries of the total integrand by computing the direct product of those irreducible representations. A direct product is an abstract way to obtain the symmetries of the product of two functions $f$ and $g$: if $h = f g$, the symmetries of $h$ may be obtained by performing the direct product of the representations of $f$ and $g$. That is, $\Gamma_h = \Gamma_f \otimes \Gamma_g$, where $\Gamma_h$ is the representation of $h$ in some point group. Since the fields have been classified already in terms of their irreducible representations, we write the irreducible representation of $\psi_{m,n}^0$ as $\Gamma_{m,n}$ and simply refer to Figs.~\ref{fig3} or~\ref{fig4}. Then, we must determine the irreducible representation of $V$, or $\Gamma_V$, which can be achieved following a process described below. We finally write the direct product as $\Gamma_{integrand} = \Gamma_m \otimes \Gamma_V \otimes \Gamma_n$.

A necessary condition for this integral to be non-vanishing is that this direct product must contain a component that transforms as a constant: the sinusoidal components do not contribute upon integration over a unit cell. Since a constant is fully symmetric (that is, it transforms as $\Gamma_1$, which is the highest symmetry irreducible representation in every point group), this condition is identical to saying that $\Gamma_{integrand}$ must contain $\Gamma_1$. Note that this condition is necessary, but not sufficient. For instance, a cosine transforms as $\Gamma_1$ about the origin, but integrates to zero over a period. We can therefore say that $\left \langle(\psi_m^0)^*V\psi_n^0\right \rangle$ is non-vanishing \textit{\textit{only if}} $\Gamma_m \otimes \Gamma_V \otimes \Gamma_n = \Gamma_1 + ...$.

A direct product is easily calculated by referring to the direct product table of the relevant point group (see Appendix \ref{A}, Fig.~\ref{Group2}). A notable feature of these tables is that two irreducible representations $\Gamma_i$ and $\Gamma_j$ satisfying $\Gamma_i \otimes \Gamma_j = \Gamma_1 + ...$, also satisfy $\Gamma_i = \Gamma_j$. Consequently, we can reframe the condition on the integrand, $\Gamma_m \otimes \Gamma_V \otimes \Gamma_n = \Gamma_1 + ...$ to be $\Gamma_m = \Gamma_V \otimes \Gamma_n$. In other words, a field $\psi_m^0$ contributes to the perturbed field $\psi_n^1$ \textit{only if} $\Gamma_m = \Gamma_V \otimes \Gamma_n$. Since $\psi_n^1$ will transform as the components comprising it (that is, the $\psi_m^0$ with non-vanishing integrals), we finally arrive at the conclusion that $\Gamma_{\psi_n^1} = \Gamma_V \otimes \Gamma_n$. Since the index $n$ refers to any particular mode of interest, we may drop it: 
\begin{equation}
\Gamma_{\psi^1} = \Gamma_V \otimes \Gamma_{\psi^0}.\label{Gpsi1}
\end{equation}
That is, the first order perturbed field profile transforms as the direct product of the irreducible representations of the perturbation operator and the unperturbed field profile in question.

Understanding the symmetries of the perturbed portion of the wavefunction allows us to simplify the free-space coupling condition, Eq. (\ref{g5}):
\begin{equation}
\gamma_e \propto \iint_A c_j\partial_i\psi \,dx\,dy = \iint_A c_j\partial_i(\psi^0 + \psi^1) \,dx\,dy \label{g6}
\end{equation}
Since $c_j$ is a function of in-plane permittivity distribution of a perturbed lattice, it is natural to expect that it has a portion that transforms like $H^0$, which we call $c_{H^0}$, and a portion that transforms like $V$, which we call $c_V$. We can write these portions explicitly to first order using the binomial approximation. Taking $c_2 = c_{H^0} + c_V$, for instance, the unperturbed portion is written
\begin{equation}
c_{H^0} = \frac{2\eta_0}{i\beta}\frac{1}{1 - \varepsilon_r^0(x,y)}, \label{ch0} 
\end{equation}
and the perturbed portion is written
\begin{equation}
c_{V} = \frac{\delta\varepsilon_r(x,y)}{1 - \varepsilon_r^0(x,y)}c_{H^0}, \label{cv}
\end{equation}
where $\varepsilon_r(x,y) = \varepsilon_r^0(x,y) + \delta\varepsilon_r(x,y)$ is decomposed into the unperturbed portion, $\varepsilon_r^0(x,y)$, and perturbed portion, $\delta\varepsilon_r(x,y)$, of the relative permittivity. Since $\varepsilon_r^0(x,y)$ transforms as $\Gamma_1$ by construction of the unperturbed lattices, it is evident that $c_{H^0}$ transforms as $\Gamma_1$: for an even function $f(x)$, the function $1/(1-f(x))$ is also even. Decomposing the factors in $c_V$, it is clear that it transforms as $\Gamma_V$ because $\delta\varepsilon_r(x,y)$ transforms as $\Gamma_V$ by definition of the perturbation, and the remaining factors in $c_V$ transform as $\Gamma_1$ (which acts as the identity in direct products). A similar argument reveals the equivalent result for $c_1$.

We therefore write $\gamma_e$ as the sum of four terms:
\begin{eqnarray}
\gamma_e \propto \iint_A c_{H^0}\partial_i\psi^0 \,dx\,dy + \iint_A c_{H^0}\partial_i\psi^1 \,dx\,dy \nonumber \\*
+\iint_A c_V\partial_i\psi^0 \,dx\,dy + \iint_A c_V\partial_i\psi^1 \,dx\,dy. \label{g7}
\end{eqnarray}
The first term vanishes for symmetry-protected BICs. As described above, $\Gamma_{H^0} = \Gamma_1$, and so, using Eq. (\ref{Gpsi1}), the second term’s integrand transforms as $\Gamma_1 \otimes \Gamma_{\partial_i} \otimes \Gamma_V \otimes \Gamma_{\psi^0} = \Gamma_{\partial_i} \otimes \Gamma_V \otimes \Gamma_{\psi^0}$. The third term’s integrand straightforwardly transforms as $\Gamma_V \otimes \Gamma_{\partial_i} \otimes \Gamma_{\psi^0}$, identical to the second term (inspection of Appendix \ref{A}, Fig.~\ref{Group2} shows that the direct products in question commute). The fourth term vanishes to first-order, because it is the product of two factors of the perturbation. We are therefore left with two generally non-vanishing terms whose integrands transform identically.

\begin{figure}
\includegraphics[width=1\columnwidth]{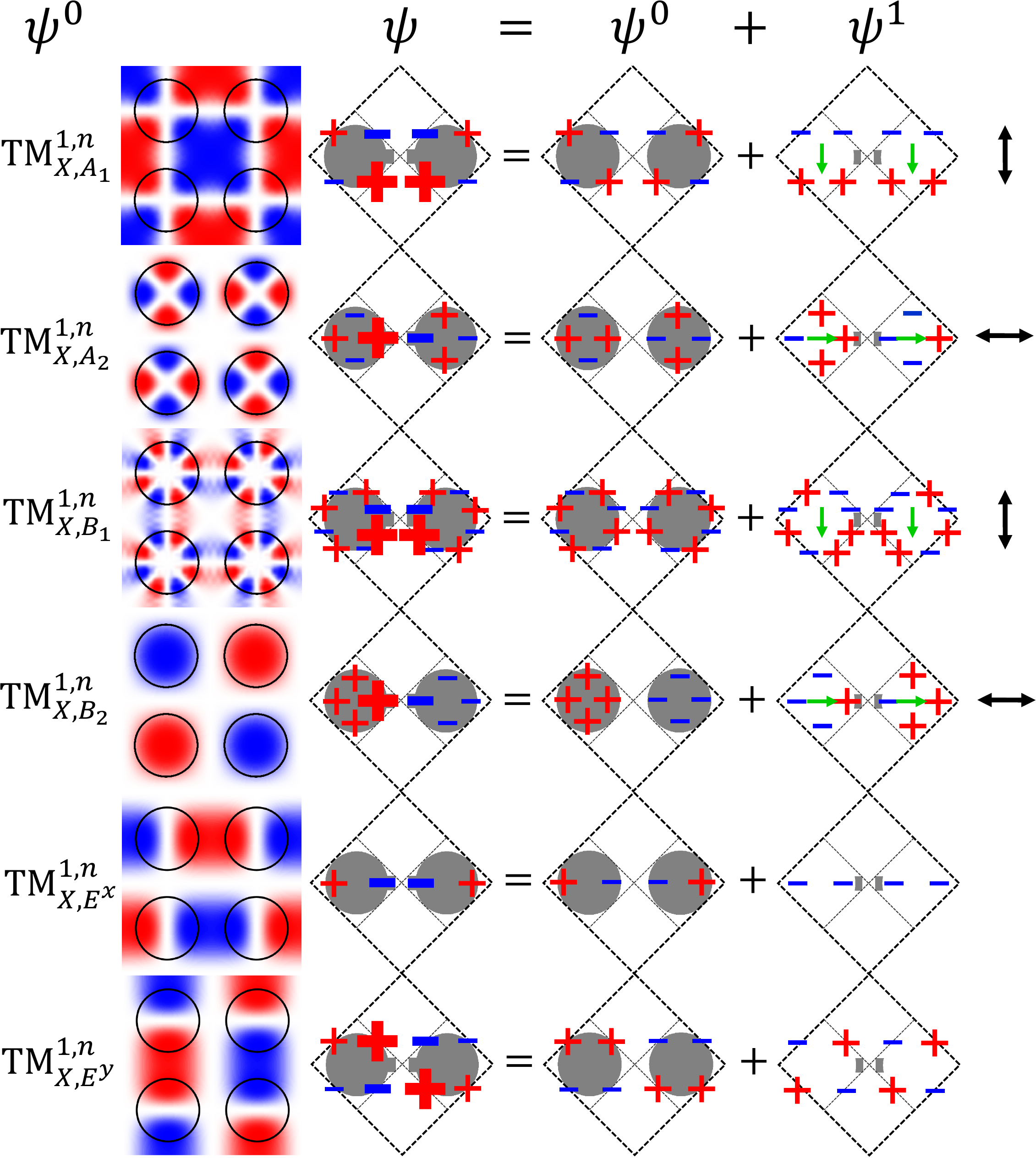}
\caption{\label{fig7}Graphical derivation of the selection rules for the $X$ point modes in the $Sq_X$ lattice belonging to the $cmm$ space group. The modes are shown in their unperturbed form as calculated by the planewave expansion method. Then, they are schematically drawn as perturbed by the perturbation and decomposed into the unperturbed portion and perturbed portion. The green arrows represent the gradient and predict coupling to a free space planewave excitation if a net dipole moment is present. The black arrows represent the corresponding free space polarization each mode couples to.}
\end{figure}

\begin{figure}
\includegraphics[width=1\columnwidth]{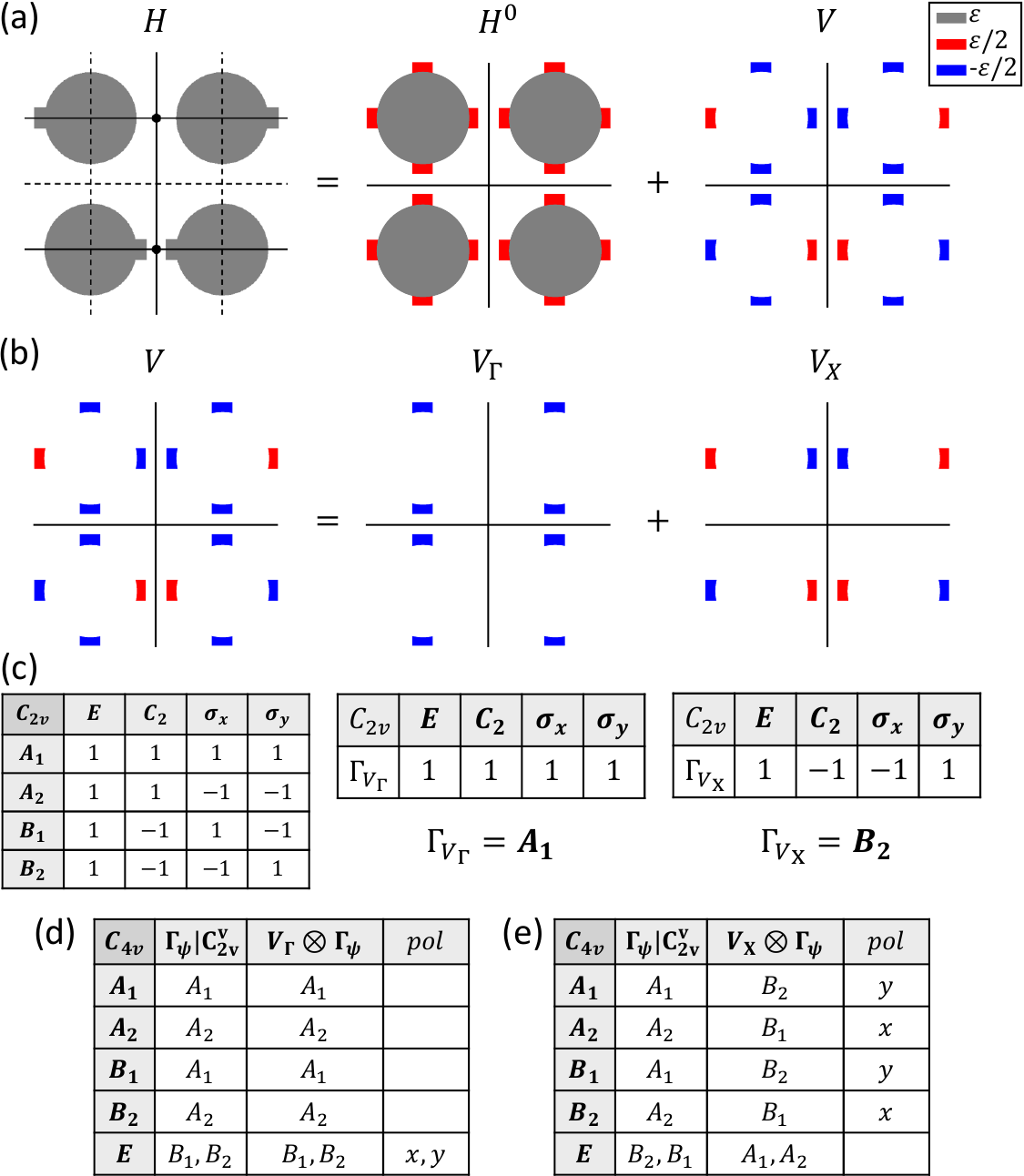}
\caption{\label{fig8}Group Theory derivation of the selection rules for the $X$ point modes in the $Sq_X$ lattice belonging to the $cmm$ space group. (a) Graphical depiction (excluding the background permittivity for simplicity) of the decomposition such that $H = H^0 + V$ for the target degenerated $Sq_X$ lattice. (b) Further decomposing $V$ into portions with different periods. (c) Determining the irreducible representation of each component of $V$ in $C_{2v}^v$ (character table reproduced for reference). (d,e) Worksheet depicting the process of deriving the selection rules. The first column shows the irreducible representations of the target modes; the second column shows the degenerated irreducible representations of those modes; the third column is the direct product of the perturbation operator and each mode; and the fourth column marks the free space polarizations matching the direct product in column 3.}
\end{figure}

As before, $\gamma_e$ is non-vanishing \textit{only if} the integrand has a component that transforms like $\Gamma_1$. We therefore arrive at the symmetry constrained coupling condition:
\begin{equation}
\Gamma_{\partial_i} = \Gamma_V \otimes \Gamma_{\psi^0}. \label{cc}
\end{equation}
Since a partial derivative in the $i$ direction transforms like a vector in that direction, it also transforms the same as a free space polarization $i$. The physical interpretation of the coupling condition, Eq. (\ref{cc}), then, is that the symmetries of the perturbed part of the field (i.e., $\Gamma_V \otimes \Gamma_{\psi^0}$) must match the symmetries of a free space polarization (i.e., $\Gamma_{\partial_i}$). That is to say, the perturbed field must have a net dipole moment to couple to free space. 

The coupling condition is equivalent to considering whether the integral
\begin{equation}
\gamma_V \propto \iint_A \partial_i(V\psi^0) \,dx\,dy \label{ccv}
\end{equation}
vanishes. This form justifies a convenient and insightful graphical method~\cite{kilic_controlling_2008} of determining whether $\gamma_e$ is non-zero, without directly determining $\psi^1$, which is not obvious at first glance at Eq. (\ref{psi1}). The perturbed mode can be simply drawn by altering the magnitude of the unperturbed field according to the shape and sign of the perturbation. Then, this new perturbed field is decomposed into the unperturbed portion and the perturbed portion (corresponding to $V\psi^0$). Taking the derivative amounts to treating the product $V\psi^0$ as ``charges'' and the gradient as the ``moment''; then, if there is a net dipole moment, the mode couples to the corresponding free space polarization. Figure~\ref{fig7} depicts this process for determining the selection rules of the $\text{TM}_{X,S}^{m,n}$ modes in a $Sq_X$ lattice with a $cmm$ space group (the same used in Fig.~\ref{fig1}). The polarization depicted corresponds to the out-of-plane field component. That is, if $\psi^0$ is a TE (TM) mode, the polarization depicted describes the magnetic (electric) polarization of free space that couples to $\psi$. Figure~\ref{fig7} therefore correctly predicts the polarization dependence seen in Fig.~\ref{fig1} for $\text{TE}_{X,B_2}^{1,1}$.

A more expedient method to generate the selection rules, however, is to determine the irreducible representations present in $V$ and then employ the direct product tables (see Appendix \ref{A}, Fig.~\ref{Group2}(a) and~\ref{Group2}(b)) to immediately write the selection rules for all modes present at the $\Gamma$ point of the perturbed lattice. This is done by (1) finding the point group in common among $V$ and $\psi$, (2) writing the irreducible representations of each factor in that point group, and then (3) determining if the direct product $\Gamma_V \otimes \Gamma_\psi$ matches the irreducible representation of a free space polarization (which are reported for each relevant point group in Appendix \ref{A}, Fig.~\ref{Group2}(c)). 

The irreducible representations of $V$ can be found by conventional Group Theory methods if required, but are generally apparent by inspection. Figure~\ref{fig8} depicts the decomposition of $V$ for the same space group as Fig.~\ref{fig7}. The process is simplified by properly choosing $H^0$ such that $V$ transforms as simply as possible. For instance, $H^0$ is written as a circle with permittivity $\varepsilon$ shadowing a square cross oriented in the $x,y$ directions with permittivity $\varepsilon/2$, as shown in Fig.~\ref{fig8}(a). It is then clear to see that the $V$ depicted obtains $H$ upon addition of $H^0$. 

Next, $V$ can be decomposed into two portions, one (called $V_\Gamma$) with the periodicity of the unperturbed lattice, and one (called $V_X$, here) with the periodicity of the perturbed lattice (Fig.~\ref{fig8}(b)). Importantly, Eq.~\ref{cc} refers only to functions as they exist in the unperturbed FBZ, in which modes characterized by the $X$ point are orthogonal to modes characterized by the $\Gamma$ point. Consequently, $V_\Gamma$ contributes only to the modes at the $\Gamma$ point in the unperturbed lattice, while $V_X$ contributes only to the modes at the $X$ point in the unperturbed lattice. The point group of both $V_\Gamma$ and $V_X$ is the same as the point group of the space group, $C_{2v}^v$. Referring to the character table of $C_{2v}^v$ (Fig.~\ref{fig8}(c) or Appendix \ref{A}, Fig.~\ref{Group1}(a)), it is readily apparent that $V_\Gamma$ transforms as $A_1$ and $V_X$ transforms as $B_2$.

\begin{figure*}
\includegraphics[width=1.4\columnwidth]{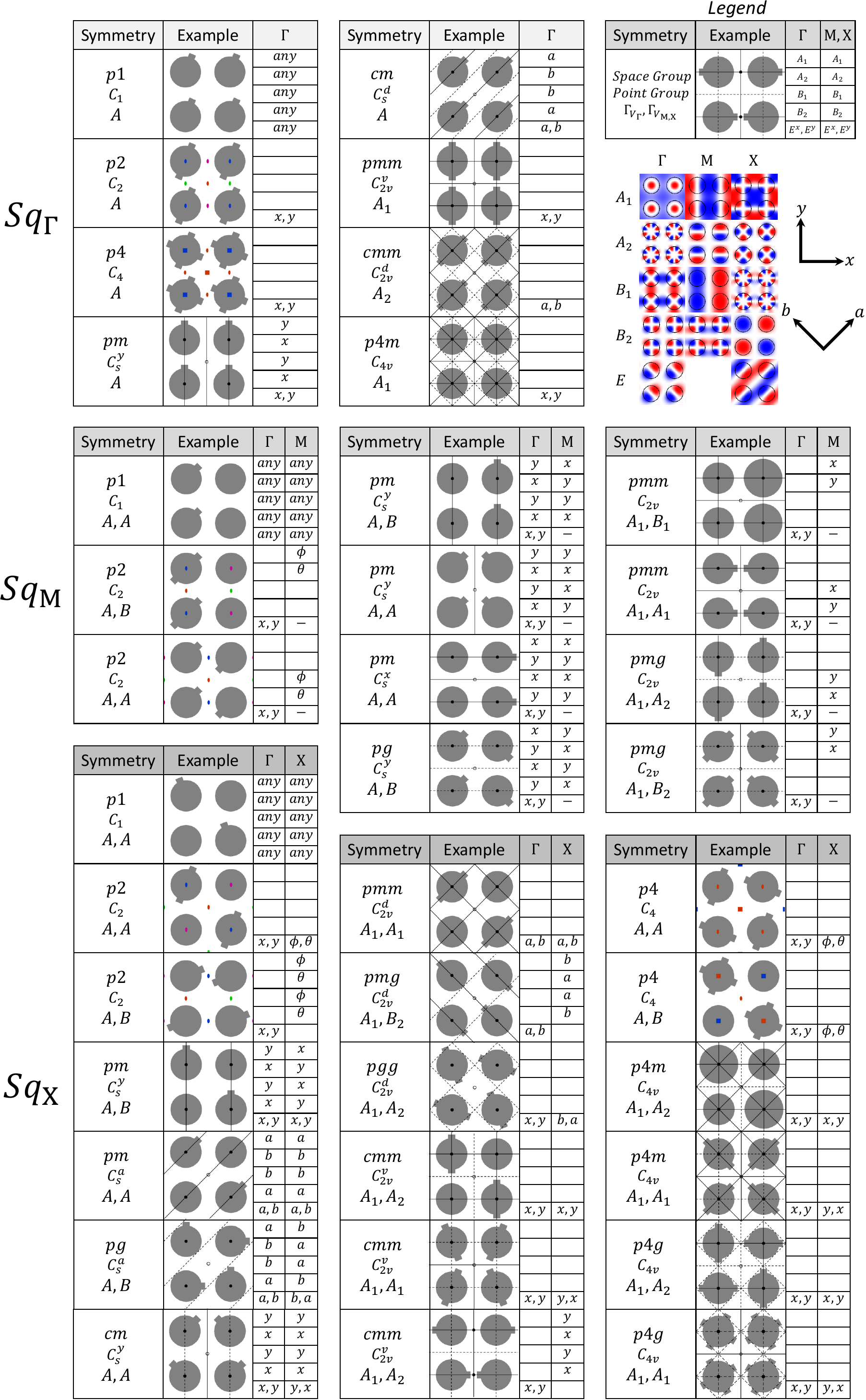}
\caption{\label{fig9}Selection rules catalog of the three square lattice families ($Sq_\Gamma$, $Sq_M$, and $Sq_X$). As depicted by the legend (upper right), along the symmetry columns each entry specifies the space group, the point group used to describe $V$, and the irreducible representations of the two components of $V$ (the first having periodicity of the unperturbed lattice, and the second of the perturbed lattice). The example column depicts an example perturbed unit cell matching the specifications in the symmetry column. Colored squares and ovals denote points with four-fold and two-fold rotational symmetry, respectively. The remaining columns are labeled $\Gamma$, $M$, or $X$ and report the selection rules for each high symmetry mode at the corresponding position in the unperturbed FBZ. Rows are indexed by the legend, and correspond to different irreducible representations present at the point in the FBZ labeled by the column (example TM modes of each row and column are shown in the legend for reference). Entries in these columns and rows refer to the free-space polarizations that excite the corresponding modes due to the perturbation and are defined by the given axes. The polarization pertains to the free space electric field for TM modes and the magnetic field for TE modes. A blank entry signifies a forbidden mode excitation (the mode remains a symmetry-protected BIC), and a dash indicates that the $M$-point does not have a fifth irreducible representation. }
\end{figure*}

\begin{figure*}
\includegraphics[width=1.5\columnwidth]{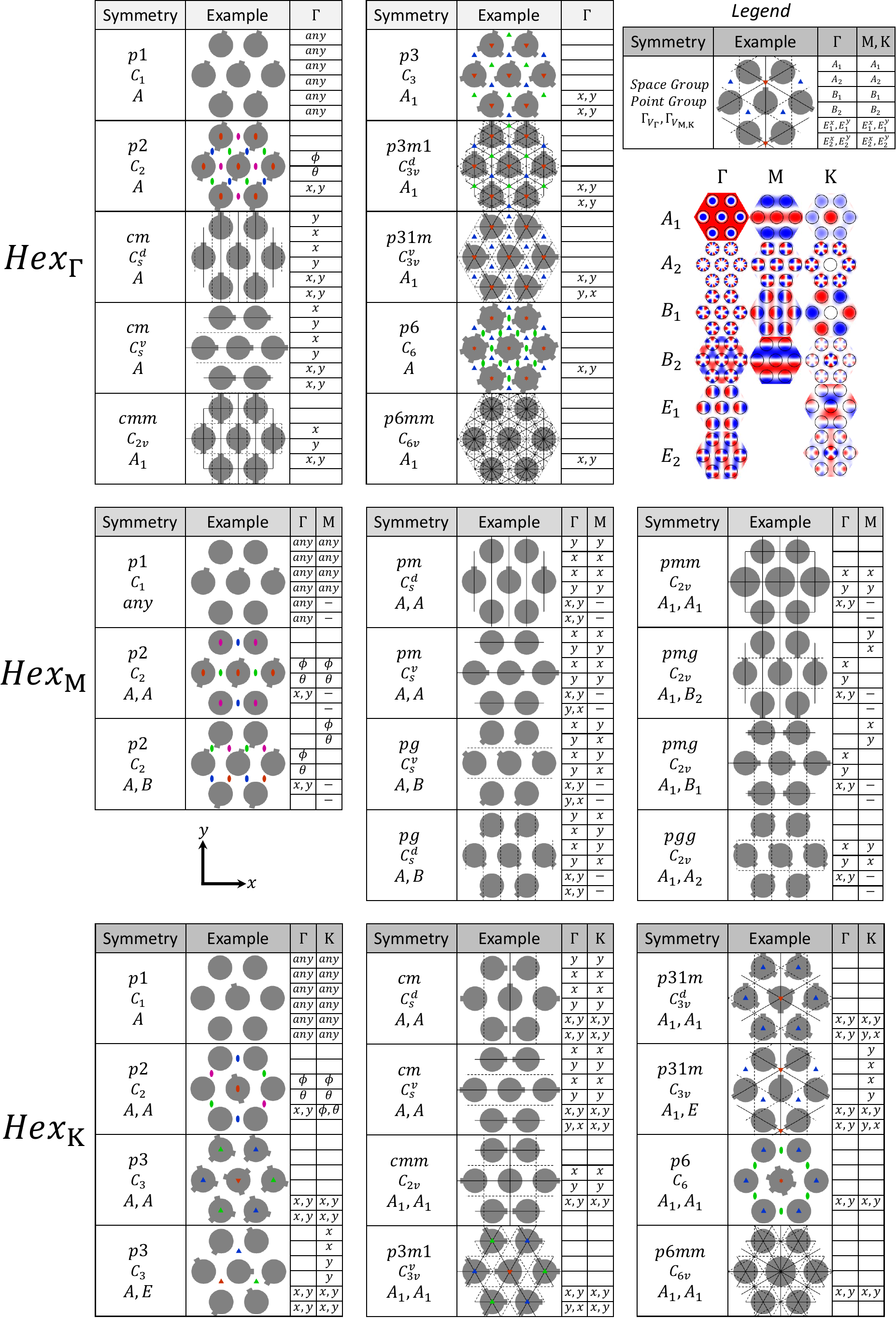}
\caption{\label{fig10}Selection rules catalog of the three hexagonal lattice families ($Hex_\Gamma$, $Hex_M$, and $Hex_K$). As depicted by the legend (upper right), along the symmetry columns each entry specifies the space group, the point group used to describe $V$, and the irreducible representations of the two components of $V$ (the first having periodicity of the unperturbed lattice, and the second of the perturbed lattice). The example column depicts an example perturbed unit cell matching the specifications in the symmetry column. Colored stars, triangles, and ovals denote points with six-fold, three-fold, and two-fold rotational symmetry, respectively. The remaining columns are labeled $\Gamma$, $M$, or $K$ aand report the selection rules for each high symmetry mode at the corresponding position in the unperturbed FBZ. Rows are indexed by the legend, and correspond to different irreducible representations present at the point in the FBZ labeled by the column (example TM modes of each row and column are shown in the legend for reference). Entries in these columns and rows refer to the free-space polarizations that excite the corresponding modes due to the perturbation and are defined by the given axes. The polarization pertains to the free space electric field for TM modes and the magnetic field for TE modes. A blank entry signifies a forbidden mode excitation (the mode remains a symmetry-protected BIC), and a dash indicates that the $M$-point does not have a fifth or sixth irreducible representation.  }
\end{figure*}

Finally, the coupling constraint (Eq. (\ref{cc})) is evaluated. However, since the modes of interest are defined in a higher group than that of $V$, we must first determine how they degenerate into the lower group. This can be done by referring to the symmetry degeneration tables (see Appendix \ref{A}, Fig.~\ref{Group1}(c)). Then, the direct products $\Gamma_{V_\Gamma} \otimes \Gamma_\psi$ are taken with reference to the direct product table for the point group $C_{2v}^v$. Since $x, y$ polarized planewaves transforms as $B_1, B_2$ in $C_{2v}^v$, the modes for which the product $\Gamma_V \otimes \Gamma_\psi = B_1, B_2$ couple to $x, y$ polarization, respectively. Figures~\ref{fig8}(d) and~\ref{fig8}(e) show a worksheet of this process. It bears repeating that this polarization corresponds to that of the out-of-plane field component. For example, if $\psi$ is a $H_z$ mode, then $x$ refers to the $H_x$ component of the free space planewave, corresponding to $y$ polarized light as conventionally defined by the electric field. The resulting selection rules are in agreement those derived using the graphical method in Fig.~\ref{fig7}, but derivation required a single diagram to decompose $V$ instead of one for each mode, and straightforwardly gave the selection rules for the $\Gamma$ point modes as well (for which the graphical method would require another set of diagrams).

The method detailed throughout this section may be summarized as follows. First, the unique space groups compatible with each lattice type are determined by exhaustion. Second, for each of these space groups, the perturbation is split into $V_\Gamma$ and $V_L$, where $L$ is the high symmetry point of the unperturbed reciprocal lattice that is folded to the $\Gamma$ point. Third, the irreducible representation of each portion of the perturbation is determined. Fourth, the coupling condition, Eq. (\ref{cc}), is evaluated for each mode, using $V_\Gamma$ for $\psi_{\Gamma,S}^{m,n}$ and $V_L$ for $\psi_{L,S}^{m,n}$; the matching polarization (if any) is marked down by reference to Fig.~\ref{Group2}(c).

With this method, a catalog for each of the six lattice types is generated, for each of the compatible degenerated space groups described above. The catalogs are given in Fig.~\ref{fig9} for square lattices, and in Fig.~\ref{fig10} for hexagonal lattices. An entry of the catalog lists the space group by name, the point group used to describe $V$, the irreducible representations of $V_\Gamma$ and $V_L$, an example visualization of the degenerated lattice (using the ``keyhole'' motif~\cite{kilic_controlling_2008}), and the selection rules for all the modes present at the $\Gamma$ point in the perturbed lattice. 

Note that the selection rules for the two-fold cyclic space group, $p2$, in the catalog are specified by some angle, $\phi$ or $\theta$, which are ill-defined relative to the lattices' axes: the polarization angle must be numerically determined, will generally change with the magnitude of the perturbation, and may differ between TM and TE modes of the same symmetry. However, for small perturbations, two modes controlled by the same $p2$ perturbation and specified by $\phi$ will be excited by the same polarization angle $\phi$; $\theta$ denotes the angle orthogonal to $\phi$. Note that the Group Theory approach in Fig.~\ref{fig8} can only say that \textit{some} polarization couples to the mode, but cannot specify $\phi$ and $\theta$; for this, the diagrammatic approach in Fig.~\ref{fig7} is used. The selection rules for $p1$ are ill-defined in a similar way: some polarization couples with a direction unconstrained by group theory, and are therefore specified as $any$. The remaining cyclic space groups, $p3$, $p4$, and $p6$, only allow access to degenerate modes in a polarization independent manner, and so for simplicity are specified as $x, y$. 

\section{Discussions and Applications}
\label{D}

The process described above lays out the derivation of the selection rules for two-dimensional PCSs with in-plane perturbations applied. The resulting catalog, split into Figs.~\ref{fig9} and~\ref{fig10}, contains a great amount of information and warrants further discussion and exploration. In particular, a few unique features present in the catalog readily motivate device applications not possible in the simpler one-dimensional PCSs. 

For instance, due to the two-dimensional nature of the device, the band structure can be optimized in both in-plane directions, allowing for full optimization of the band structure and thereby optimally compact devices. One-dimensional structures (e.g., devices composed of rectangular grating fingers, invariant in one in-plane direction) can be understood as a special case of a subsection of the $Sq_M$ lattice, but with limited to no control over the behavior along the direction of the grating fingers. Additionally, the higher in-plane symmetry of two-dimensional structures means the presence of degenerate $E$-type modes (``partner'' modes with identical eigenfrequencies that couple to orthogonal polarizations), which do not exist in one-dimensional structures. This allows for compact, polarization independent devices such as filters and modulators to be designed. The manipulation of degenerate modes is therefore of considerable technological interest. Last, we note a parent-child relationship between higher and lower order space groups within the catalog, and find that child space groups constructed by successively adding distinct parent space groups result in optical control with independent degrees of freedom introduced by the parent space groups. We show that this principle enables controlling a large number of parameters characterizing an optical spectrum, well surpassing the state-of-the-art.

With these considerations in mind, Sec.~\ref{D1} discusses the degenerate modes that exist in monatomic and multi-atomic PCSs, providing a comprehensive set of options for polarization-independent devices using the catalog. Section~\ref{D2} details a device application motivated by this discussion, demonstrating that the degenerate fundamental modes $Hex_K$, as controlled by three successive perturbations, are suitable for Terahertz generation via four-wave mixing. As a second application of successive perturbations apparent from studying the entries of the catalog, Sec.~\ref{D3} shows the potential for a PCS on a stretchable substrate to enable mechanically tunable optical lifetimes. Finally, Sec.~\ref{D4} reports the discovery of a geometric phase associated with circularly polarized light coupling into a $p2$ space group of the $Sq_M$ lattice, controllable simultaneously with the Q-factor of the resonance. We then show how spatially varying this geometric phase enables a novel class of photonic devices in which the outgoing Fano resonant wavefront is spatially tailored while the non-resonant light is left unaffected.

\subsection{Degenerate Modes}
\label{D1}

We first consider the nature of degenerate modes. The degenerate modes generally transform as partners of a degenerate irreducible representation (e.g., $E$ in $C_{4v}$), which are written $E^x$ and $E^y$ corresponding to their dipole moments. Because of this dipole moment, the $E^x$ and $E^y$ modes in the unperturbed $Sq_\Gamma$ lattice generally couple to free space (and the $E_1$ modes couple to free space in the unperturbed $Hex_\Gamma$ lattice). In other words, the integral 
\begin{equation}
\iint_A c_{H^0}\partial_i\psi^0 \,dx\,dy, \label{cch}
\end{equation}
which has an integrand that transforms as $\Gamma_1 \otimes \Gamma_{\partial_i} \otimes \Gamma_{\psi^0}$, is non-vanishing for $E$ modes because $\Gamma_{\partial_i} = E$ in $C_{4v}$ (likewise, $\Gamma_{\partial_i} = E_1$ in $C_{6v}$). The coupling can numerically vanish for certain combinations of angle, polarization, and optical materials, but since these are reasons unrelated to the symmetry arguments above, they are accidental BICs; all modes other than the $\psi_{\Gamma,E}^{m,n}$ for $Sq_\Gamma$ and $\psi_{\Gamma,E_1}^{m,n}$ for $Hex_\Gamma$ in the mode classification are symmetry-protected BICs. For the modes already coupled to free space in the unperturbed lattice, the only significant impact a perturbation has is to split the degeneracy upon symmetry degeneration (for instance, perturbing a lattice with $C_{4v}$ down to $C_{2v}$). In other words, if the lattice is made structurally birefringent, the $E^x$ and $E^y$ modes will degenerate into irreducible representations in a lower order point group with different eigenfrequencies, but the coupling rate to free space will generally be changed to a negligible degree. For this reason, Figs.~\ref{fig9}~and~\ref{fig10} simply label the corresponding entries $x, y$.

Of more interest here are the degenerate symmetry-protected BICs. The $M$ point modes of either square or hexagonal lattices have no such degenerate modes because the $C_{2v}$ point group has no degenerate irreducible representation (a rectangle is not identical in the $x, y$ directions). However, the $\psi_{\Gamma,E_2}^{m,n}$ for $Hex_\Gamma$, $\psi_{X,E}^{m,n}$ for $Sq_X$, and $\psi_{K,E_1}^{m,n}$ and $\psi_{K,E_2}^{m,n}$ for $Hex_K$ are degenerate symmetry-protected BICs. Therefore, a polarization insensitive filter or modulator must use one of these lattices in order to utilize the advantages of quasi-BICs (that is, a Q-factor controllable by Eq. (\ref{Qdelt}) independent of the band structure). We consider the degenerate modes in each of these three lattices in turn.

The $Hex_\Gamma$ supports the $E_2$ modes, which are uncoupled to free space in the absence of a perturbation ($E_1 \otimes E_2 = B_1 + B_2 + E_1$, which does not contain $\Gamma_1$ in $C_{6v}$) and are degenerate BICs. Reference to the catalog shows that reducing the symmetry to $C_{3v}$ or lower may allow free-space coupling to these modes. A polarization independent filter or modulator with Q-factor following Eq. (\ref{Qdelt}) could be made utilizing the $E_2$ modes of a $Hex_\Gamma$ lattice according to either the $p31m$ or $p3m1$ entry of the catalog.

The $Sq_X$ lattice supports degenerate modes that are bound in the unperturbed lattice. Upon perturbation, they are brought to the $\Gamma$ point, allowing coupling to free space at normal incidence. Several space groups in the $Sq_X$ lattice leave these modes uncoupled in the continuum, making them BICs, while most others allow coupling, making them quasi-BICs. The space groups with $C_{4v}$ and $C_4$ leave the eigenfrequencies degenerate, while lower order symmetry groups introduce birefringent behavior. Therefore, a polarization independent filter or modulator with Q-factor following Eq. (\ref{Qdelt}) may be made utilizing the $E$ modes of a $Sq_X$ lattice according to any of the $p4m$, $p4g$, or $p4$ entries of the catalog as reference.

\begin{figure}
\includegraphics[width=0.95\columnwidth]{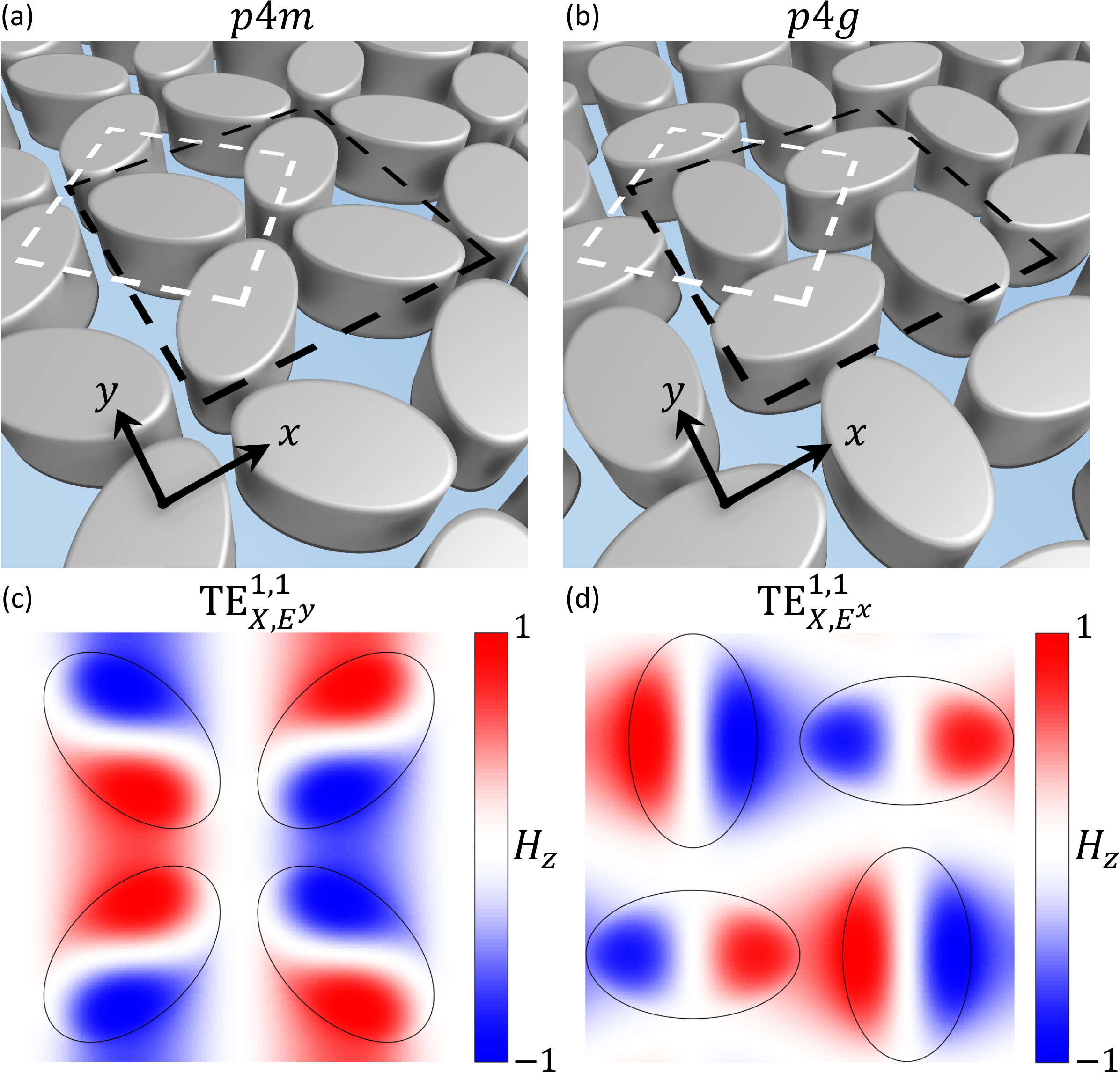}
\caption{\label{fig11}Mode ``twisting'' in a $Sq_X$ lattice. (a) Example $p4m$ lattice. (b) Example $p4g$ lattice. The white dashed boxes in (a) and (b) denote the perturbed unit cell, and the black dashed boxes denote the plotting area in (c) and (d). (c) Normalized field distribution for the $\text{TE}_{X,E^y}^{1,1}$ partner excited by a planewave with magnetic polarization in the $x$ direction. (d) Normalized field distribution for the $\text{TE}_{X,E^x}^{1,1}$ partner excited by the same polarization in (c).}
\end{figure}

An interesting feature of the catalog is the prediction of coupling of the $E^x$ partner of the $E$ modes of a $Sq_X$ lattice to either $x$ or $y$ polarized light (equivalently, $y$ polarized light may couple into either the $E^x$ or $E^y$ partner). Compare, for instance, the $p4m$ with $V_X = A_1$ (Fig.~\ref{fig11}(a)) and $p4g$ with $V_X = A_2$ (Fig.~\ref{fig11}(b)) space groups in the $Sq_X$ lattice. Figures~\ref{fig11}(c) and~\ref{fig11}(d) depict, for the $p4m$ and $p4g$ cases respectively, the field profiles calculated by full-wave simulations at the frequency of the $\text{TE}_{X,E}^{1,1}$ modes excited by $y$ polarized light (magnetically $x$ polarized light). The former shows that the magnetically $x$ polarized planewave couples to the $E$ mode with the apparent dipole in the $y$ direction (that is, the $E^y$ partner, as defined in Fig.~\ref{fig7}), while for the latter it couples to the $E$ mode with the apparent dipole in the $x$ direction (that is, the $E^x$ partner). This mode ``twisting'' is written in the catalog by writing $y, x$ for $p4m$, in contrast to the entry of $x, y$ for $p4g$, and it is easily predicted by tracking how the partners degenerate to $C_{2v}$ in the worksheet of Fig.~\ref{fig8}, or by using the diagrammatic approach in Fig.~\ref{fig7}. This is a phenomenon that does not occur in the more often studied $E$ or $E_1$ degenerate modes of the $Sq_\Gamma$ or $Hex_\Gamma$ lattices, and is thus representative of the larger range of behaviors present in multi-atomic lattices. Especially notable is that the dependence of the polarization angle of the incident light on the in-plane orientation of elliptical structures suggests that a geometric phase is associated with this coupling. This will be explored in Section~\ref{D4}.

Next, the $Hex_K$ lattice is a special case, having degenerate modes of two distinct types. The unit cell is a trimer, and therefore contains three times the number of modes as the unperturbed lattice. As described in Sec.~\ref{B2}, two sets of modes (corresponding to $K_1$ and $K_2$ points in the FBZ in Fig.~\ref{fig5}) are folded to the $\Gamma$ point. Because of the symmetry of the unperturbed lattice, a mode in one of the sets has a counterpart in the other set with equal eigenfrequency. These pairs mix at the perturbed $\Gamma$ point, producing a final set of modes describable in $C_{6v}$. The lattice therefore supports $E_1$ and $E_2$ modes newly brought to the $\Gamma$ point by the perturbation, analogous to the $E_1$ and $E_2$ modes in the $Hex_\Gamma$ lattice.

However, inspection of the band diagram for the $Hex_K$ lattice near the $\Gamma$ point in Fig.~\ref{fig5} (bottom right) shows many more degeneracies than explainable by the $E_1$ and $E_2$ modes alone. All of the newly folded modes, in fact, are degenerate, despite the mode classification scheme in Fig.~\ref{fig2} predicting the presence of modes not describable by $E$ irreducible representations. In particular, the fundamental modes are degenerate (last band diagram in Fig.~\ref{fig5}), but have irreducible representations $A_1$ and $B_1$. Although visibly quite different (see the $K^{(1)}$ column of Fig.~\ref{fig4}), and having distinct symmetries in $C_{6v}$, they are nonetheless identical in eigenfrequency. This degeneracy is born of the trimerization of the lattice: a pair of modes with the same eigenfrequency are superposed upon translation to the $\Gamma$ point, and can be superposed either in phase or out of phase, producing a pair of distinct modes with identical eigenfrequency.

One consequence of this is that in the $p31m$ $Hex_K$ lattice with point group $C_{3v}^d$, the $A_1$ and $B_1$ modes form a degenerate pair that together correspond to a spectral feature that is polarization insensitive (a similar behavior is seen in the $p3$ lattice where $V_K$ transforms as a partner of the $E$ irreducible representation: the fourth entry in the $Hex_K$ lattice). This reveals another way to consider this degeneracy: the $A_1$ and $B_1$ modes are partners of the $E$ irreducible representation of the $C_{3v}^d$ point group, defined about the $\kappa$ point in the real space lattice (as defined in the $Hex_\Gamma$ lattice in Fig. ~\ref{fig5}). Because the $\gamma$ point has the full symmetries of $C_{6v}$, this description of the modes about the $\kappa$ point misses relevant symmetries; nevertheless, the ability to describe them in $C_{3v}^d$ as partners of the $E$ modes means their eigenfrequencies must be identical. Upon further symmetry degeneration (for instance, to $cmm$, or $cm$ in the $Hex_K$ lattice), the modes behave differently, splitting in both eigenfrequency and polarization dependence. This behavior is unique to the $Hex_K$ lattice in Figs.~\ref{fig9} and \ref{fig10} because it is the only lattice with more than two atoms per unit cell. Higher order lattices, such as those shown in Fig.~\ref{B1}, may exhibit similar behavior.

\subsection{Application: Terahertz Generation}
\label{D2}

\begin{figure*}
\includegraphics[width=1.9\columnwidth]{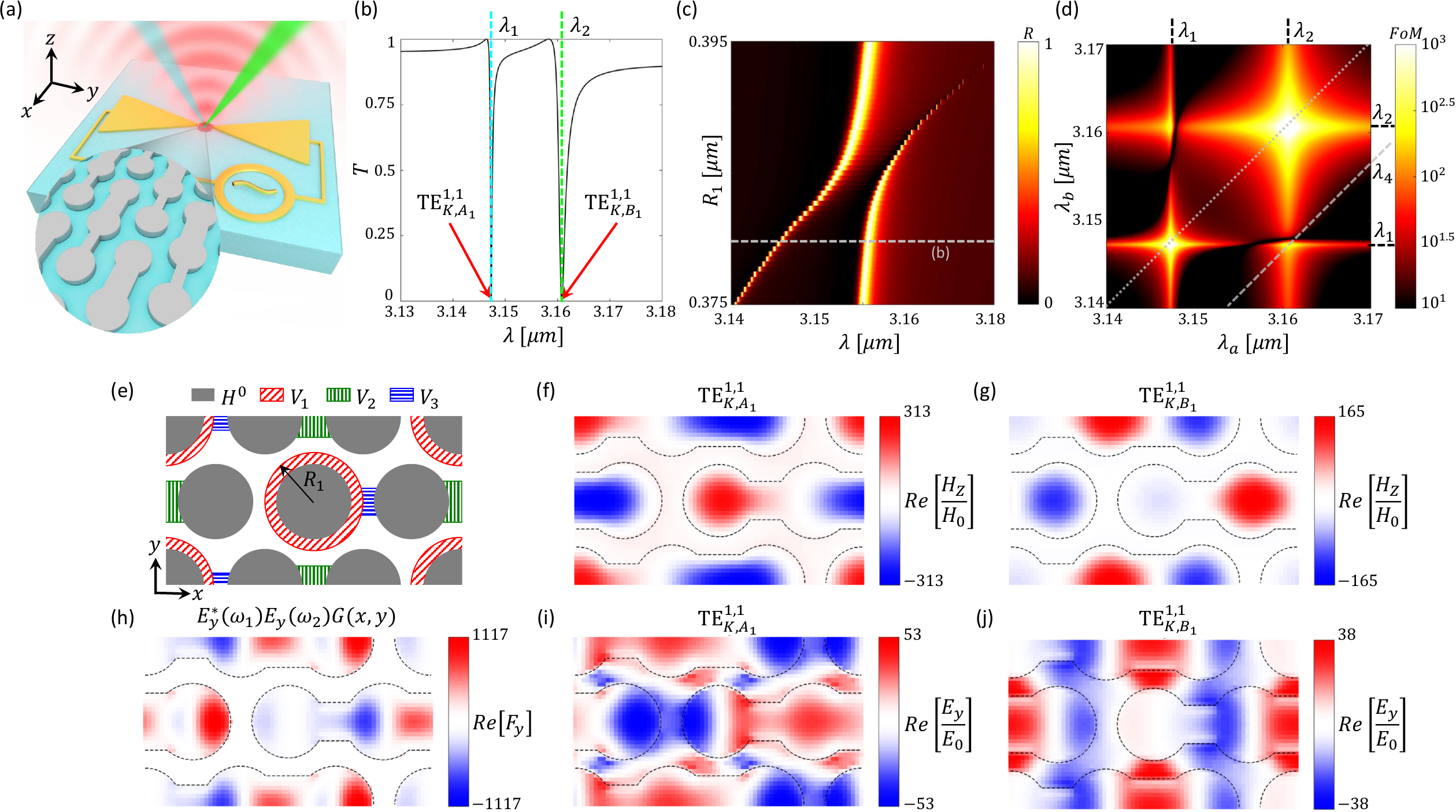}
\caption{\label{fig12}Terahertz generation with four-wave mixing. (a) Schematic of device excited by two (narrowband) near-infrared pump lasers (shown in cyan and green), producing Terahertz radiation (shown radiating in red). (b) Transmittance near $\text{TE}_{K,A_1}^{1,1}$ and $\text{TE}_{K,B_1}^{1,1}$ modes, which are degenerate in the unperturbed lattice. (c) Map of reflectance, $R$, showing control of the frequency spacing of the two resonant modes by altering the radius of the center pillar in (e). (d) Map of the figure of merit, $FoM$, with a dashed contour for $1/\lambda_4 = 1/\lambda_1 - 1/\lambda_2$ shown. The $FoM$ is maximized along this contour for Terahertz generation at the coordinates ($\lambda_1$, $\lambda_2$), corresponding to enhancement due to both resonances. (e) Successive perturbations to the unperturbed hexagonal lattice. $V_1$ controls the frequency spacing between the two resonant modes in (b), $V_2$ controls the Q-factor of the $B_1$ mode, and $V_3$ controls the Q-factors of both the $A_1$ and $B_1$ modes. (f,g) Magnetic field profiles for the $\text{TE}_{K,A_1}^{1,1}$ and $\text{TE}_{K,B_1}^{1,1}$ modes normalized to the magnetic field of the incident planewave, $H_0$. (h) Visualization of the integrand in the $FoM$, calculated using (i,j) the $E_y$ components of $\text{TE}_{K,A_1}^{1,1}$ and $\text{TE}_{K,B_1}^{1,1}$ normalized to the electric field of the incident planewave, $E_0$. }
\end{figure*}

We explore the degenerate fundamental modes of the $Hex_K$ lattice to aid in generating Terahertz (THz) frequencies through nonlinear processes enhanced by optical resonances. Sketching the design of such a device is a useful exercise to demonstrate the utility and an example use of the catalog. Figure~\ref{fig12}(a) depicts a schematic of the device, with a $Hex_K$ lattice made of Silicon pillars in the gap of a bowtie antenna resonant to a THz frequency. Figure~\ref{fig12}(b) shows an example spectrum of the PCS portion of this device, showing two closely spaced resonances at $\lambda_1 = 3.147 \mu m$ and $\lambda_2 = 3.161 \mu m$, both excited by $y$ polarized light and associated with the split degeneracy of the $\text{TE}_{K,A_1}^{1,1}$ and $\text{TE}_{K,B_1}^{1,1}$ modes. If optical power is normally incident at pump wavelengths $\lambda_a$ and $\lambda_b$ such that $\lambda_a = \lambda_1$ and $\lambda_b = \lambda_2$, and a low-frequency bias (corresponding to a radiofrequency with wavelength $\lambda_3$) is electrically applied across the antenna, four-wave mixing will produce photons at a THz wavelength with improved efficiency (compared to a bulk material) due to the enhanced light-matter interactions from the PCS and antenna resonances. The case shown in Fig.~\ref{fig12}(b) corresponds to $\lambda_4 = 711 \mu m$, but Fig. ~\ref{fig12}(c) shows that $\lambda_4$ can be easily tuned by the radius of the central pillar, $R_1$. Figure~\ref{fig12}(d) confirms that the figure of merit (defined below) is indeed maximum at $\lambda_4$ when the pump photons have wavelengths of $\lambda_1, \lambda_2$ ($1/\lambda_3 = 0$ for simplicity, here; it may generally be used to finely and actively tune $\lambda_4$). 

A key advantage of using these degenerate modes is the unique robustness of the control of both the spectral spacing and linewidths of the resonances. Since the modes are degenerate in the unperturbed lattice, they are necessarily closely spaced in a weakly perturbed lattice. Then, by controlling the radius of one of the pillars, the frequency spacing can be finely tuned. The spectral map in Fig. ~\ref{fig12}(c) shows the impact of tuning the radius of the central pillar, $R_1$, depicting a classic anti-crossing behavior~\cite{fan_temporal_2003} as the resonance spacing changes. This utilization of degenerate modes offers a considerably more robust control of closely spaced resonances compared to relying on controlling two unrelated resonances by tuning geometric parameters: fine tuning the separation of two unrelated resonances is highly sensitive to fabrication errors, while the split degeneracy here is guaranteed by symmetry.

Notably, $\lambda_4$ (or the spacing of the resonances) can be tuned largely independently of the linewidths of the resonances. This is easily understood by considering the portion of the perturbation, $V_1$, that corresponds to changing $R_1$. Depicted in Fig.~\ref{fig12}(e), a lattice perturbed by $V_1$ alone produces a $Hex_K$ lattice with the $p6mm$ space group; reference to the catalog (Fig. ~\ref{fig10}) reveals that no coupling to the target modes is introduced by this perturbation. Tuning the radius of the central pillar therefore does not affect the coupling to first order. Then, the addition of $V_2$ degenerates the space group to $cmm$, which couples the $\text{TE}_{K,B_1}^{1,1}$ mode to the magnetic $x$ polarization, but not the $\text{TE}_{K,A_1}^{1,1}$ mode. Finally, the addition of $V_3$ creates a lattice with the $cm$ space group with the $C_{s}^v$ point group, allowing coupling the $\text{TE}_{K,A_1}^{1,1}$ mode to the magnetic $x$ polarization. Notably, if the other $cm$ space group (with point group $C_{s}^d$ in the third table of Fig.~\ref{fig10}) were used, the two resonances would be cross-polarized. Tuning these three portions of the perturbation therefore allows independent tuning of each of the linewidths and the spacing of the two resonances in either a co-polarized or cross-polarized fashion. The co-polarization of the two previously degenerate resonances is unique to the $Hex_K$ lattice in the catalog, as the $E_1$ and $E_2$ modes of the hexagonal lattices (and $E$ modes in the square lattices) are only accessible in a cross-polarized fashion. Co-polarized split degenerate states are a unique feature of lattices with more than two atoms (such as the lattices containing four atoms seen in, for instance, Fig.~\ref{B1}). In the present application, the freedom to have the pump wavelengths be co-polarized allows a single pulse (with bandwidth spanning the two resonances) as the pump. 

To complete the demonstration of the advantages of the $Hex_K$ lattice for Terahertz generation, we compute a simple figure of merit related to the efficiency of this conversion (see, for instances, Refs.~\cite{lin_cavity-enhanced_2016,lee_giant_2014}): 
\begin{equation}
FoM = \left \lvert \iint \chi^{(3)}(x,y)E^*(\omega_1)E(\omega_2)E^*(\omega_3)E^*(\omega_4) \,dx \,dy \right \rvert \label{fom1}
\end{equation}
where the bounds of integration are over the entire device and $\chi^{(3)}(x,y)$ is the spatially dependent third-order nonlinear susceptibility and the electric fields are normalized to the corresponding incident fields.
Given the scale difference of $\lambda_3$ to a unit cell (i.e., $\lambda_3^2 \gg A$), a reasonable approximation to this integral is that $E^* (\omega_3 ) = F_3$ and $E^* (\omega_4) = F_4$ are constants equal to the electric field enhancement due to the bowtie antenna. We may then integrate over a unit cell:
\begin{equation}
FoM = \left \lvert \chi_{Si}^{(3)}F_3^*F_4^* \iint_A F_y(x,y) \,dx \,dy \right \rvert \label{fom}
\end{equation}
where $F_y = E_y^* (\omega_1)E_y(\omega_2)G(x,y)$ and $G(x,y) = 1$ where there is Silicon and is $0$ where there is vacuum. That is, the figure of merit is proportional to the overlap integral of the two pumps within the Silicon portion of a unit cell. 

The integrand may be calculated from the mode profiles taken from full-wave simulations of unit cell of the device. The $H_z$ component of the modes for the spectrum in Fig.~\ref{fig12}(b) are shown in Figs.~\ref{fig12}(f) and~\ref{fig12}(g), corresponding to the choice in Fig.~\ref{fig12}(c) of $R_1 = 0.38 \mu m$. The integrand of Eq. (\ref{fom}) is shown in Fig.~\ref{fig12}(h), as calculated from the $E_y$ components of the modes (shown in Figs.~\ref{fig12}(i) and~\ref{fig12}(j)) and the refractive index profile of the device. The numerical value for this case is $FoM/\chi_{Si}^{(3)} \approx 114\lvert F_3\rvert \lvert F_4\rvert$, meaning that for a modest enhancement of $\left \lvert F_3\right \rvert = \left \lvert F_4\right \rvert \approx 10$ by the bowtie antenna we will have a total enhancement of $FoM/\chi_{Si}^{(3)} \approx 10^4$ in efficiency. Figure~\ref{fig12} therefore demonstrates a platform to produce THz light from two infrared pumps taking advantage of large electric field enhancement at every frequency involved in the four-wave mixing process. The phase matching condition is guaranteed by the subwavelength scale of the device in the vertical direction, the resonance spacing is robustly controlled by the radius of the central pillar, and the resonant linewidths can be tuned largely independently by the successive degeneration from $C_{6v}$ to $C_{2v}$ to $C_{s}$. We note that there is some partial cancellation upon integration of the integrand of Eq. (\ref{fom}), but not complete cancellation. Future work could optimize the perturbations chosen such that this cancellation is minimized.

\subsection{Application: Mechanically Tunable Optical Lifetimes}
\label{D3}

\begin{figure}
\includegraphics[width=1\columnwidth]{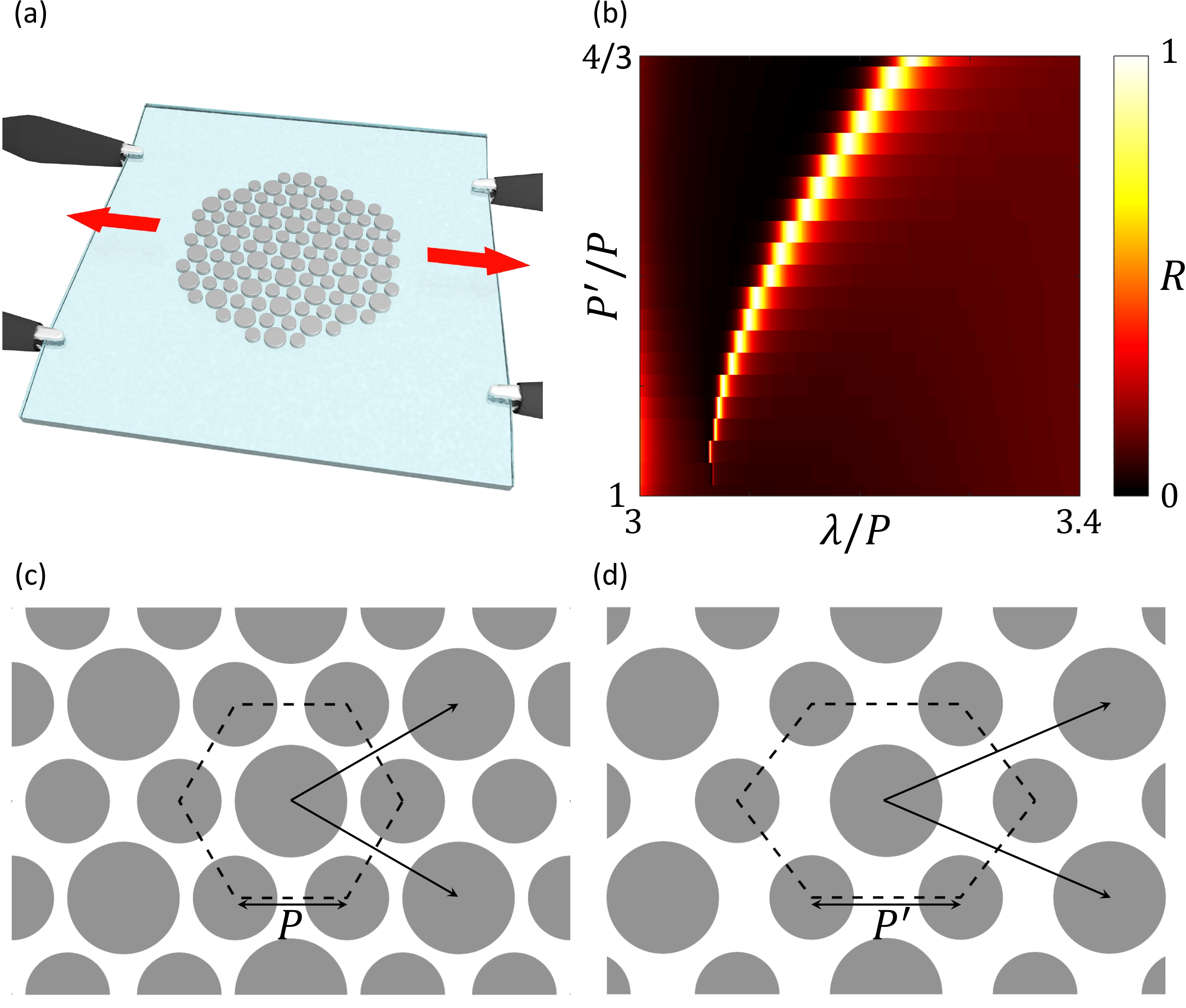}
\caption{\label{fig13}Mechanically tunable optical lifetime. (a) Schematic of a $Hex_K$ lattice with a $p6mm$ space group on a stretchable substrate. (b) Full-wave simulations mapping the spectral reflectance, $R$, near the wavelength, $\lambda$, of the $\text{TE}_{K,B_1}^{1,1}$ mode for various degrees of stretching. (c) The unstretched lattice, with spacing $P$. (d) Stretched lattice, with spacing $P'$ along the horizontal. Single sided arrows denote the lattice vectors in (c) and (d).}
\end{figure}

Next, we remark on a type of periodic perturbation achieved by stretching or shearing a high symmetry lattice. Since the symmetry of the lattice is reduced, the symmetry-protected BICs may be excited. However, the condition on coupling (that is, Eq. (\ref{cc})) still applies, and therefore this class of perturbation follows the same selection rules as the equivalent point group degeneration entries. For instance, by shearing an unperturbed $Sq_\Gamma$ lattice’s unit cell from a square into a rhombus, the space group is reduced to $p1$ and any mode at the $\Gamma$ point may now couple to free space with a strength related to the degree of shear. However, the polarization direction of the coupled planewave will be ill-defined in general, changing, for instance, with the degree of shear. (Recall that it is for this reason that the $p1$ entries are all specified as $any$, because no general comment can be made.)

Of more interest is stretching along a high symmetry axis, affording well-defined selection rules. This has been explored in Ref.~\cite{cui_dynamic_2012} for plasmonic heptamers arranged in a square lattice by degenerating the symmetry of the heptamer from $C_{6v}$ to $C_{2v}$ by stretching the substrate. Since the Fano resonance in the plasmonic heptamer is both (1) well-confined to a unit cell of the overall lattice, and (2) due to the coupling between plasmonic modes, analysis of the point group of the unit cell alone suffices to analyze the resonance. However, for a low loss, high Q-factor demonstration using dielectric structures, this analysis is insufficient because the coupling across unit cells of the array is integral to the presence of BICs. The catalog of selection rules derived above provides the necessary information for proper analysis in dielectric systems. 

Inspection of the catalog of the square lattices (Fig.~\ref{fig9}) reveals that square lattices afford no interesting cases: the only impact of a lattice deformation along a high symmetry axis is to split degeneracies, not introduce any new coupling. This is not true for the hexagonal lattices, however. Figure~\ref{fig13}(a) shows a $Hex_K$ lattice with $p6mm$ space group (one pillar of the trimer has a larger radius than the others) on a stretchable substrate. In the unstretched case, the lattice has $C_{6v}$ symmetry, and the selection rules forbid coupling to any but the $E_1$ modes at normal incidence. However, inspection of the $cmm$ space group reveals that degeneration from $C_{6v}$ to $C_{2v}$ enables coupling to the $B_1$ and $B_2$ modes. Stretching the $Hex_K$ lattice with $p6mm$ space group along the $x$ axis also degenerates the point group from $C_{6v}$ to $C_{2v}$, and so ought to enable coupling to those modes to a degree controlled by the strength of the lattice deformation. Figure~\ref{fig13}(b) depicts confirmation of this prediction via full-wave simulations near the $\text{TM}_{K,B_1}^{1,1}$ mode, showing redshift and a changing Q-factor as a function of deformation. Inspection of the $Hex_K$ catalog (or an analogous case in the $Hex_\Gamma$ lattice) therefore enables a low-loss dielectric-based flexible device platform with mechanically tunable resonant lifetime. 

\subsection{Nonlocal Metasurfaces}
\label{D4}

The preceding applications demonstrated the utility of the catalog to guide device design using successive perturbations. In other words, the key to their design came from understanding how the final, lower order space group was constructed from the higher order space groups, which together form a parent-child relationship. In this section, we supplement the approach in the accompanying Letter~\cite{overvig_multifunctional_2020} in which we use this principle to demonstrate how particular $p2$ space groups may be constructed from two parent space groups. We elucidate a geometric phase that is a consequence of the parent space groups exhibiting mode ``twisting'' of the sort shown in Fig.~\ref{fig11}, and how if this geometric phase is spatially varied, we may realize devices with anomalous reflection and refraction only on resonance. 

\begin{figure}
\includegraphics[width=1\columnwidth]{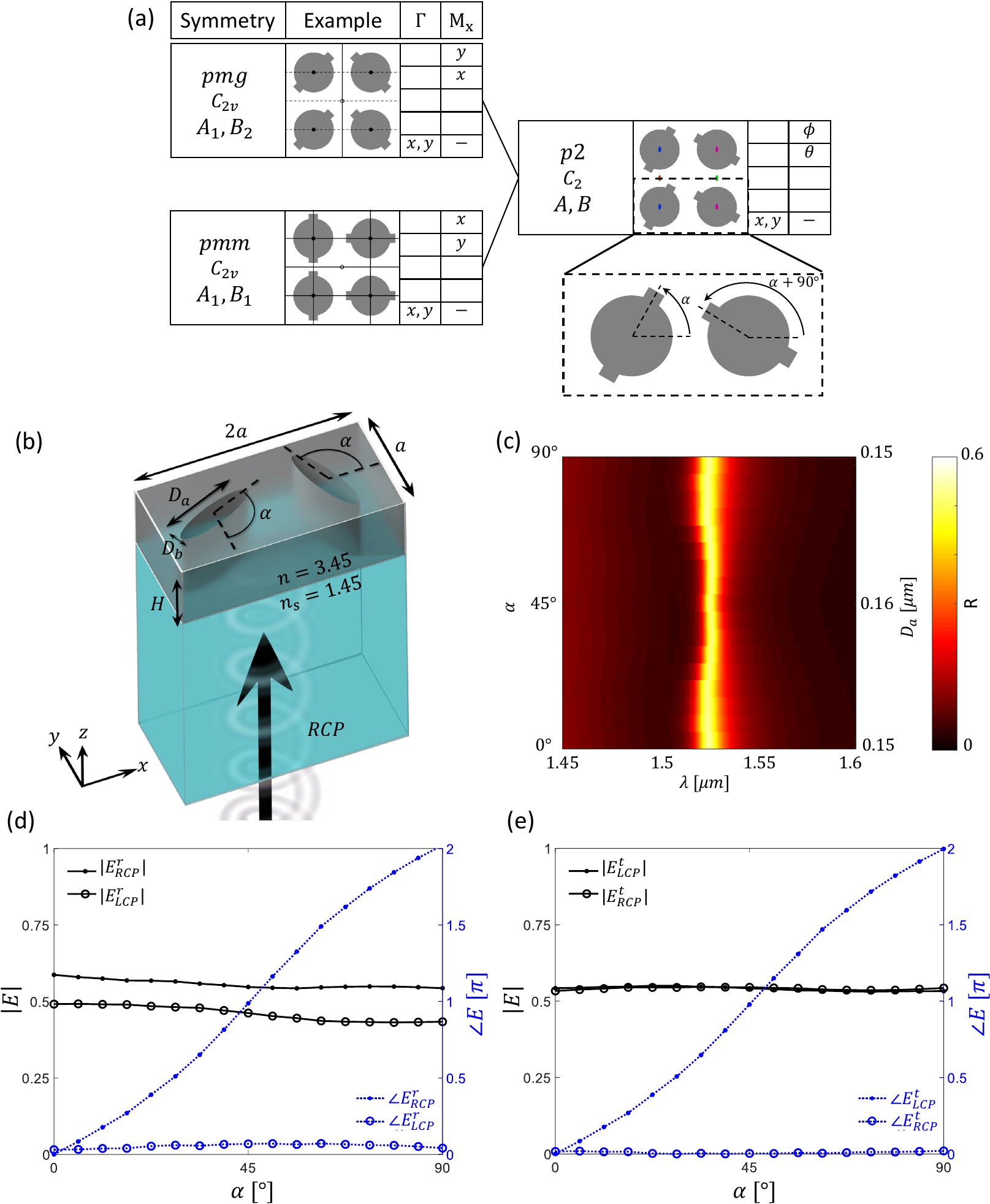}
\caption{\label{fig14}Meta-units that introduce two factors of the geometric phase. (a) Hierarchical relationship between a $p2$ (child) space group and two higher order (parent) space groups in the $Sq_M$ lattice. (b) Schematic of a meta-unit composed of ellipses etched into a slab of Silicon, excited by right hand circularly polarized (RCP) light incident from the substrate. (c) Reflectance map for a meta-unit library constructed by varying $\alpha$ and $D_a$ to keep a constant resonant frequency. (d) Amplitude and phase responses of the LCP and RCP components of the reflected light on resonance. (e) Amplitude and phase responses of the LCP and RCP components of the transmitted light on resonance.}
\end{figure}

We begin by focusing on the relationship between three space groups in the $Sq_M$ lattice, shown in Fig.~\ref{fig14}(a). Two parent space groups ($pmg$ and $pmm$) are shown on the left. These two space groups share no symmetries in common except two-fold rotations at the center of the Silicon pillars. Consequently, if the perturbations are added successively, the child space group will retain only these two-fold rotations, resulting in the $p2$ space group shown on the right in Fig.~\ref{fig14}(a). This example of a parent-child relationship between higher order parents and lower order children is very general; the full hierarchy for the $Sq_M$ and $Sq_X$ lattice is reported in Fig.~\ref{figC1}. (Note: the hierarchy particular to the multi-wavelength metasurface introduced in the accompanying Letter~\cite{overvig_multifunctional_2020} is reported in Fig.~\ref{figC2})

Examining Fig.\ref{fig14}(a) shows that the $pmm$ parent space group allows for coupling to a polarization angle $\phi = 0\degree$ from the $x$ axis (that is, $x$ polarization), while the $pmg$ parent space group allows for coupling to a polarization angle $\phi = 90\degree$ from the $x$ axis (that is, $y$ polarization). As seen in the inset of Fig.\ref{fig14}(a), the child space group may be parameterized by an orientation angle, $\alpha$, that yields the $pmm$ parent space group when $\alpha = 0\degree$, and the $pmg$ parent space group when $\alpha = 45\degree$. In other words, as $\alpha$ varies continuously from $0\degree$ to $45\degree$, the corresponding polarization angle must vary from $0\degree$ to $90\degree$. The linear interpolation of this behavior is
\begin{equation}
\phi = 2\alpha. \label{geo} 
\end{equation}
This form is highly reminiscent of the well-known geometric phase, $2\alpha$, which is introduced when the handedness of circularly polarization is flipped while light is scattered by an anisotropic scatterer oriented along the $\alpha$ direction. This similarity suggests studying a $p2$ space group under circularly polarized illumination, as shown in Fig.~\ref{fig14}(b) for RCP light.

In particular, upon studying the phase of circularly polarized light exiting the device on both the reflection side and transmission side, we find that this system imparts a geometric phase to light exiting with the converted handedness (for RCP incidence, this is LCP in transmission, and RCP in reflection) that is twice the conventional geometric phase. As in conventional dichroic optical elements (e.g., a plasmonic bar antenna), two projections of the polarization basis are required to analyze the outgoing light, one from coupling into the element and the second from coupling out. We will consider each in turn. 

First, only the component of free-space light that is linearly polarized in the $\phi$ direction completley couples to the mode. This light, which constitutes half of the power of the RCP incident light, is resonantly reflected, while orthogonally polarized light (at an angle $\theta=\phi+90\degree$) is transmitted. Decomposing the incident RCP light into two linearly polarized components, the $\phi$ component carries a phase $\Phi^r_1=\phi$, and the $\theta$ component carries a phase $\Phi^t_1=\phi+90\degree$. Since the $\phi$ direction is defined by the orientation angle $\alpha$ by Eq.~\ref{geo}, the  resonantly reflected light is therefore associated with a phase $\Phi^r_1 =  2\alpha$ and the orthogonally polarized transmitted light has a phase $\Phi^t_1 = 2\alpha+90\degree$.

Second, the output light on resonance is linearly polarized, and can be decomposed into its constituent LCP and RCP components. These components have a geometric phase $\Phi^r_2=\mp\phi$ in reflection and $\Phi^t_2=\pm\theta$ in transmission, where the first sign corresponds to LCP, and the second corresponds to RCP. We can finally determine the total phases $\Phi = \Phi_1 + \Phi_2$ of the LCP and RCP components in reflection and transmission:

\begin{subequations}
\begin{align}
\Phi^r_{LCP}&= 2\alpha - 2\alpha &&= 0\degree  \\
\Phi^r_{RCP}&= 2\alpha + 2\alpha &&= 4\alpha \\
\Phi^t_{LCP}&= (2\alpha + 90\degree) + (2\alpha + 90\degree) &&= 4\alpha+180\degree \\
\Phi^t_{RCP}&= (2\alpha + 90\degree) - (2\alpha + 90\degree) &&= 0\degree.
\end{align}
\end{subequations}

That is, RCP light in reflection and LCP light in transmission vary as $4\alpha$, while LCP light in reflection and RCP light in transmission are invariant to $\alpha$. The two factors of the geometric phase come from the two instances changing the basis for the polarization state: first from circular to linear (coupling into a single linear state), and second from linear to circular (decomposing into its constituent spins). In this case, the final value is twice the conventional geometric phase because the eigenpolarization (characterizing the projection bases) varies as $\phi=2\alpha$ compared to the conventional case of $\phi=\alpha$ (e.g., a  plasmonic bar antenna oriented in-plane by an angle $\alpha$). We note that the form $\phi=2\alpha$ is not a general rule for quasi-BICs; for instance, the $p2$ space group in the $Hex_K$ lattice with a cross-motif follows $\phi=-4\alpha$ for the $B_1$ mode. This results in a geometric phase that is $-8\alpha$.

\begin{figure*}
\includegraphics[width=1.9\columnwidth]{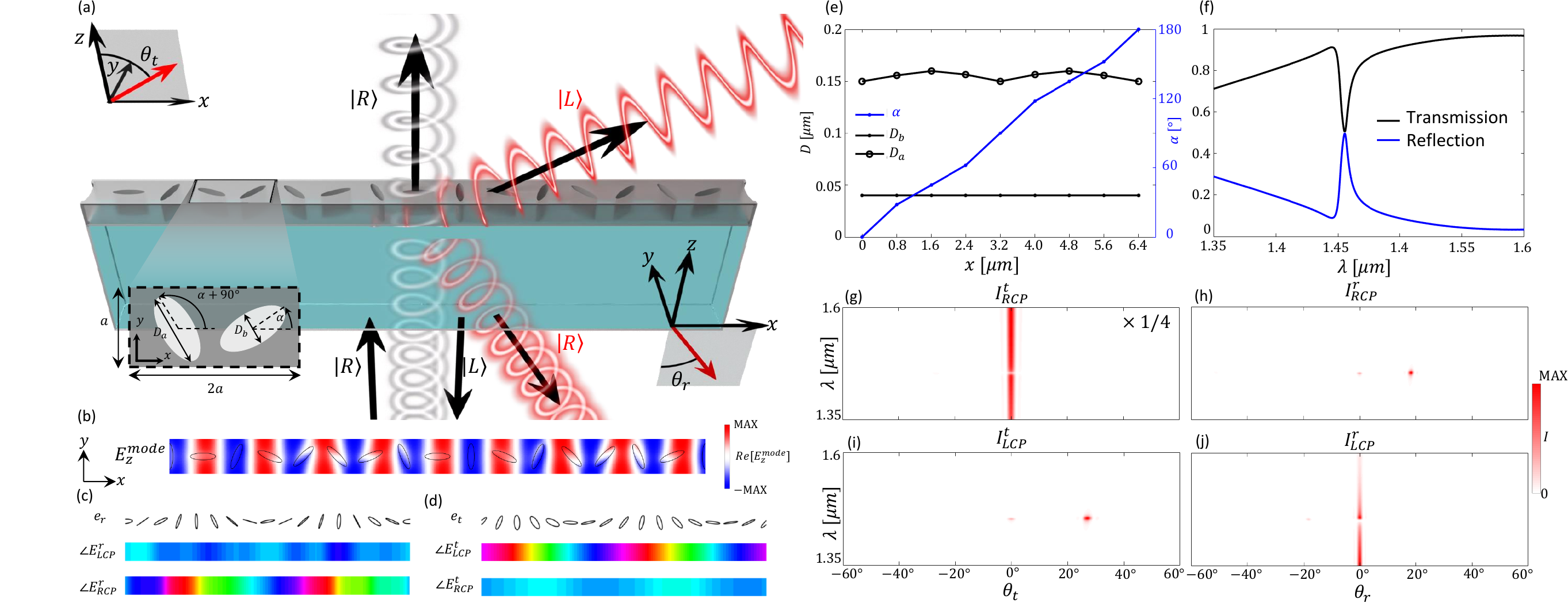}
\caption{\label{fig15}Gradient resonant metasurface. (a) Schematic depicting the device and its functionality: a thin film of Silicon on top of quartz is patterned with elliptical holes (inset shows top-view of the geometry), resonantly deflecting light with converted handedness when excited by circularly polarized light. (b) Top-view of the complex field on resonance overlaid on the gradient resonant metasurface. (c,d) Top-view of spatial distributions of the reflected and transmitted polarization states, $e_r$ and $e_t$, and phase responses of $E_{RCP}^r$, $E_{RCP}^t$, $E_{LCP}^r$, and $E_{LCP}^t$. (e) Geometrical parameters of the device in (b). (f) Transmission and reflection spectra of the device in (b). (g-j) Farfield angular and spectral intensity distributions calculated from the optical near field, such as the results in (c,d), showing deflection of light with converted handedness only on resonance.}
\end{figure*}

We next explore the physics and applications exploiting this geometric phase as a new degree of freedom. Because the geometric phase is completely controlled by $\alpha$, we may use the remaining geometric degrees of freedom of the unit cell to maintain a spatially constant resonant frequency across a device with a spatially varying geometric phase profile. In other words, as in conventional metasurface approaches, we may construct a library of geometries (``meta-units'') such that full phase coverage and constant amplitude are achieved. Then, by spatially arranging these meta-units, an output wavefront with a designer phase profile may be realized at the resonant frequency. 

To confirm this approach, we construct such a meta-unit library targeting wavelengths in the telecommunications range. A meta-unit, seen in Fig.~\ref{fig14}(b), is composed of a Silicon slab with two etched ellipses, which are identical but for a $90\degree$ rotation. The chosen thickness of the slab is $H = 250nm$, and the lattice constant is $a = 400nm$. The in-plane geometric parameters that are varied to construct the meta-unit library are the diameters along the semi-major axis, $D_a$, and semi-minor axis, $D_b$, and the orientation angle, $\alpha$. For simplicity, we keep $D_b$ constant, and vary $D_a$ and $\alpha$ so as to achieve full phase coverage with minimal shift in resonant wavelength.

A spectral map of reflectance, calculated by fullwave simulations, is shown in Fig.~\ref{fig14}(c) illustrating a near constant resonant wavelength across the meta-unit library. The amplitude and phase of the reflected (transmitted) LCP and RCP components are recorded in Fig.~\ref{fig14}(d) (Fig.~\ref{fig14}(e)) at the operating wavelength, $\lambda_{op} = 1.52 \mu m$. The amplitudes of the LCP and RCP components are approximately equal (each representing roughly one quarter of the input power) and vary little across the meta-unit library. The small inequality is due to the presence of the substrate breaking the symmetry in the out-of-plane direction. The phase of the component with converted handedness (which, for reflection is RCP, and for transmission is LCP) varies across $2\pi$ as $\alpha$ varies across $90 \degree$, and follows closely with the predicted $\Phi = 4\alpha$ dependence (see Figs.~\ref{fig14}(d,e)). 

With the meta-unit library constructed, a wavefront with a spatially shaped phase profile within a narrow bandwidth near $\lambda_{op}$ may be realized. A common function is to linearly vary the output phase so as to create anomalous reflection and refraction. We may choose either to vary the phase profile in the same direction as the dimerization (the $x$ direction in Fig.~\ref{fig14}(b)), or the orthogonal direction. We will begin with the former choice (see the accompanying Letter for the latter~\cite{overvig_multifunctional_2020}). 

Figure~\ref{fig15}(a) shows a schematic of a device deflecting the component with the converted handedness at the resonant wavelength for RCP light normally incident from the substrate side. Figure~\ref{fig15}(b) depicts the electric field on resonance calculated by fullwave simulations, overlaid with the geometry of the device (Fig.~\ref{fig15}(e)). Figure~\ref{fig15}(c,d) show the output polarization states, and phases of the LCP and RCP components at the reflection side and transmission side, respectively. In both cases, the output polarization is approximately linear across the device. As such, the phase of the signal with unconverted handedness is uniform while the phase of the signal with converted handedness varies across $4\pi$ as $\alpha$ varies over $180 \degree$.

Figure~\ref{fig15}(f) confirms that the resonance of the metasurface remains intact, despite the variance of geometry across the metasurface. However, a noticeable blueshift has occurred relative to the originally chosen $\lambda_{op}$. Nevertheless, at the resonant peak of the device, $\lambda_{dev} = 1.46\mu m$, deflection to the second diffractive order occurs for signal with the converted handedness (Figs.~\ref{fig15}(g-j)). A device with identical deflection angle is shown in the accompanying Letter~\cite{overvig_multifunctional_2020}, but with a phase gradient applied in the orthogonal direction to the dimerization direction. The blueshift is also present in that case, but significantly reduced.

\begin{figure}
\includegraphics[width=1\columnwidth]{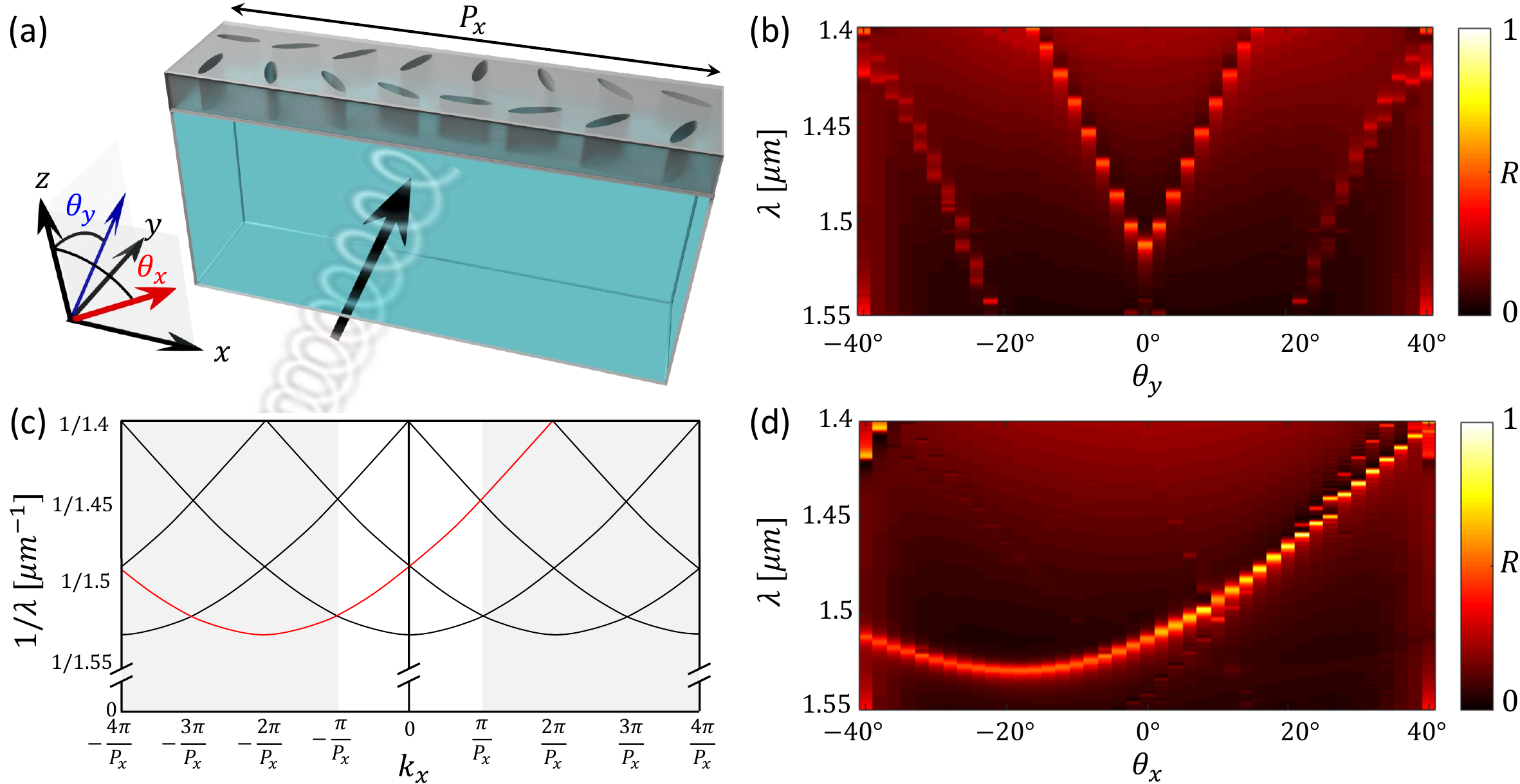}
\caption{\label{fig16}Angular dispersion of resonant metasurfaces. (a) Schematic of a resonant metasurface excited from off normal angles. (b) Reflectance map while varying $\theta_y$, showing that the resonance follows the dispersive band of the concave up mode. (c) Region of an extended zone (shaded gray) band diagram near the resonance of the device in (a) with artificial Brillouin zone folding; the red band corresponds to the band shifted by a $k$-vector equal and opposite to that introduced by the phase gradient due to coupling into the supermode (i.e., one factor of the geometric phase). (d) Reflectance map while varying $\theta_x$, showing that the resonance follows the band shifted by a factor of the geometric phase gradient. }
\end{figure}

The explanation for this blueshift comes from a unique feature of this metasurface: that the deflection of light with converted handedness is mediated by a supermode of the device. That is, unlike conventional metasurfaces, whose meta-units scatter light based on local resonances, this metasurface scatters light due to a global resonance (associated with a supermode) supported by many neighboring meta-units. To explore the physics here, we consider the dependence of a gradient resonant metasurface on the incident angle of the RCP light. Figure~\ref{fig16}(a) schematically shows a device with a spatial phase gradient in a direction orthogonal to the dimerization direction, with light incident from the substrate at a set of angles $\theta_x$ (along the phase gradient) and $\theta_y$ (along the dimerization direction). Since the deflection only occurs on resonance, the resonant frequency follows some dispersion relation (i.e., the band structure). Figure~\ref{fig16}(b) depicts the resonant frequency dispersion while varying $\theta_y$ from $-40\degree$ to $40\degree$ (corresponding, by Snell's Law, to $\pm68.7\degree$ in air). This mode is concave up, meaning that at higher in-plane momenta, a blueshift occurs.

To understand the blueshift at normal incidence, we must consider (1) the modes supported by the device and (2) the in-plane momentum of the resonant mode. First, these resonant modes exist in the device with a super-period of $P_x = 8a$ in the phase gradient direction. We therefore consider all of the supermodes supported by a device with this super-period. Because the super-period is composed of perturbed versions of the same PCS, the supermodes will be well approximated by artificial Brillouin folding corresponding to period doubling a meta-unit three times. Figure~\ref{fig16}(c) depicts such a process for the mode in question, showing the band of the unit cell (containing two ellipses) copied every integer multiple of the grating vector $k_G = 2\pi/P_x$ in the $k_x = k_0 sin(\theta_x)$ direction. The supermodes present at normal incidence are the modes at $k_x = 0$ in this diagram. 

Second, upon coupling in, there is a spatially varying geometric phase, corresponding to twice the local rotation angle of the ellipses. The derivative of this spatial phase is equivalent to a $k$-vector, 
\begin{equation}
k_{geo} = \frac{\partial \Phi}{\partial x} = k_G = \frac{2\pi}{P_x}.
\end{equation}
In other words, the resonant supermode is the resonant mode of a unperturbed lattice modulated by an in-plane wavevector in the $x$-direction. This corresponds to a supermode that is $k_{geo}$ away from the unperturbed $\Gamma$ point, $k_x = 0$ (the flat part of the band). The supermode is highlighted red in Fig.~\ref{fig16}(c). Consequently, as seen in Fig.~\ref{fig16}(d), as $\theta_x$ is varied, the resonance follows the dispersion of the band as it existed in the unperturbed lattice, shifted by $-k_{geo}$. Notably, this also means that the resonant frequencies corresponding to the pair of incident momenta $k_x=0$ and $k_x= - 2k_{geo}$ are identical. This is consistent with the requirements of reciprocity: these two momenta are the input and output momenta of the deflection process; reversing the output must yield the original input at all frequencies. 

Finally, we comment on the achievable phase gradient limited by this angular dispersion. As encapsulated by Eq.~\ref{Qbk}, the component $k$-vectors involved with the resonance must be limited according to the linewidth and angular dispersion of the resonance in order to maintain large resonance visibility. For a resonant metasurface lens or hologram shaping an incident planewave, the range of output $k$-vectors must satisfy Eq.~\ref{Qbk}. In the cylindrical  metasurface lens reported in the accompanying Letter~\cite{overvig_multifunctional_2020}, the resonance visibility is maintained despite the range of deflection angles in the $\theta_x$ direction across the device, which may be characterized by the numerical aperture, $NA$. We find that increasing the $NA$ gradually reduces the resonance visibility, but a substantial resonance visibility (a peak reflectance of $>40\%$) is still maintained at a high value of $NA = 0.7$. This is not true for a cylindrical lens focusing in the $\theta_y$ direction, where $NA<0.1$ is required to retain appreciable resonance visibility, as the angular dispersion is large in the $\theta_y$ direction compared to that in the $\theta_x$ direction (comparing Figs.~\ref{fig16}(b,d)). This is consistent with the constraint on Q-factor, band flatness, and spread in incident k-vector encapsulated by Eq.~\ref{Qbk}. In other words, it suggests that by including band structure engineering, a fully radially focusing resonant metasurface lens may be realized, and that we must generally take care to engineer the band structure of the unperturbed resonant metasurfaces before applying the perturbation, as laid out in the three-step process described in Sec.~\ref{Back}. 

Given the lack of impact on the non-resonant light waves, which may transmit with high efficiency irrespective of incident angle, we anticipate these resonantly deflecting and focusing metasurfaces to be of significant interest to augmented reality displays, which aim to superimpose a desired image on top of information transmitted through the glass originating from the external world. By further application of the principle of successive perturbations, we show in the accompanying Letter~\cite{overvig_multifunctional_2020} that the single-wavelength resonant metasurfaces may be extended to multi-wavelength devices with independently tunable phase profiles. The hierarchy of the child space group constructed from eight parents is shown in Fig.~\ref{figC2}. The eight parents represent eight degrees of freedom to spatially and spectrally shape an incident wavefront: the Q-factors and polarization angles (i.e., geometric phases) of four modes with distinct symmetries may be controlled simultaneously. Notably, these eight degrees of freedom are in addition to the degrees of freedom present in the unperturbed lattice, which may be used to control the resonant frequencies and band curvatures of the desired modes. This degree of spatial and spectral control over an optical spectrum greatly surpasses the state-of-the-art, and is readily apparent from careful study of the catalog in conjunction with the design principle of successive perturbations.

\section{Summary}
\label{Summary}

In summary, we derived the selection rules for Fano resonances due to quasi-bound states in the continuum supported by photonic crystal slabs. Targeting the high symmetry modes of both square and hexagonal lattices, we explored six lattices designed to bring each class of high symmetry mode into the continuum. We exhaustively reported the degenerated space groups due to in-plane perturbations that are compatible with these six lattices, and cataloged the selection rules in each case by applying principles of Group Theory to determine the free space polarization of the leaky portion of the perturbed modes. 

Together with band structure engineering, the principles, approach, and results outlined here provide a high-level guide to designing compact photonic crystal slabs supporting sharp resonances: devices confining light in both space and time, and manufacturable by mature fabrication technologies. Future work will be well-guided by the rational design principles considered here to reduce the search space required to optimize a compact, resonant optical device. In particular, we showed that the band structure may be engineered in the unperturbed lattice before a periodic perturbation is applied to couple the targeted mode(s) to the desired free space polarization(s). We showed, here and in the accompanying Letter~\cite{overvig_multifunctional_2020}, that in addition to the degrees of freedom present in choosing the unperturbed lattice, a series of successive perturbations may realize multifunctional control of the resonances (up to eight parameters at once). The insights of the catalog of selection rules produced by Group Theory arguments have straightforwardly motivated novel devices, such as polarization independent planar optical modulators, Terahertz generation in photonic crystal slabs with lifted degeneracies, devices with mechanically tunable optical lifetimes, and a novel class of metasurfaces that uses two factors of a geometric phase to spatially shape a resonant wavefront. We therefore believe that careful understanding and examination of the patterns and features of the selection rules represent a fruitful launching point for future efforts.

\begin{acknowledgments}
The authors thank J. Lee-Thorpe and M. Weinstein, and Andrea Al\`{u} for helpful discussions. The work was supported by the Defense Advanced Research Projects Agency (grant nos. D15AP00111 and HR0011-17-2-0017), the National Science Foundation (grant nos. ECCS-2004685 and QII-TAQS-1936359), and the Air Force Office of Scientific Research (grant no. FA9550-14-1-0389). A.C.O. acknowledges support from the NSF IGERT program (grant no. DGE-1069240). S.C.M acknowledges support from the NSF Graduate Research Fellowship Program (grant no. DGE-1644869). M.J.C. acknowledges support from USD/R\&E (The Under Secretary of Defense-Research and Engineering), National Defense Education Program (NDEP)/BA-1, Basic Research.
\end{acknowledgments}

%

\appendix
\section{Group Theory Tables}
\label{A}
\renewcommand{\thefigure}{A~\arabic{figure}}
\setcounter{figure}{0}

\begin{figure*}
\includegraphics[width=1.85\columnwidth]{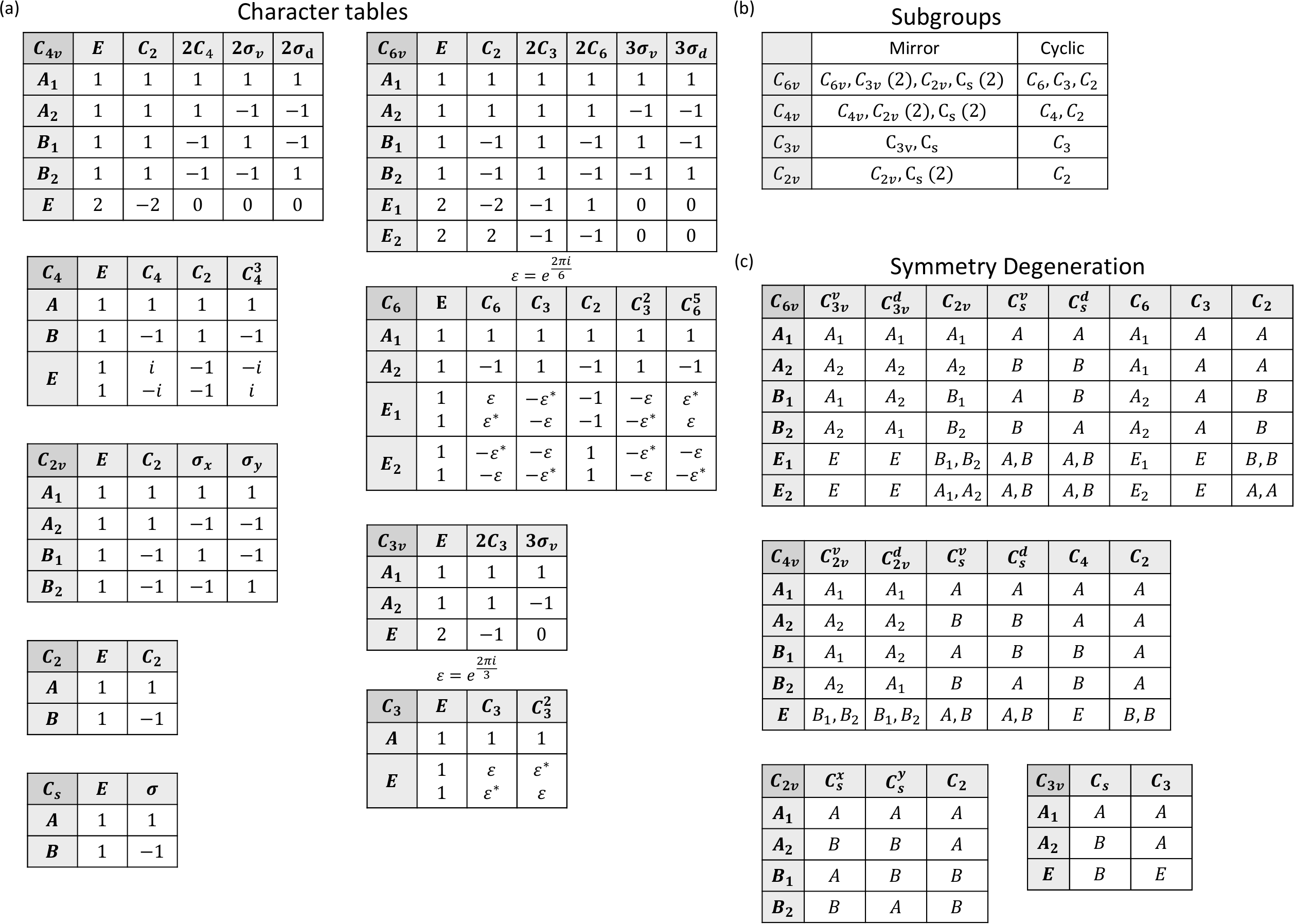}
\caption{\label{Group1}Group Theory tables for point groups. (a) The character tables of the point groups relevant to the square (left) and hexagonal (right) lattices. (b) The subgroups of the higher symmetry lattices. (c) The symmetry degeneration tables, describing how irreducible representations in higher order groups degenerate in lower order groups.}
\
\end{figure*}

For ease of reference, the character tables of all relevant point groups are reported in Fig.\ref{Group1}(a). The left column of tables in Fig.~\ref{Group1}(a) contains the point groups compatible with the square and rectangle lattices, and the right contains those compatible with hexagonal lattices. Figure~\ref{Group1}(b) summarizes the subgroups of each of the point groups shown in Fig.~\ref{Group1}(a). This prescribes the necessary components of the symmetry degeneration tables, shown in Fig.~\ref{Group1}(c), which track how higher symmetry modes (irreducible representations) degenerate into lower groups. That is, reference to the symmetry degeneration tables provides the answers to how a higher symmetry mode would be named in a lower order symmetry group (for instance, $B_2$ in $C_{4v}$ would be called $A_2$ in $C_{2v}^v$).

\begin{figure}
\includegraphics[width=1\columnwidth]{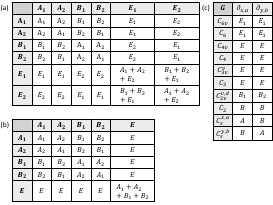}
\caption{\label{Group2}Group Theory tables for deriving the selection rules. (a) The direct product table for $C_{6v}$. (b) The direct product table for $C_{4v}$. The direct product tables for the lower order point groups are subsets of (a) and (b). (c) The irreducible representation describing the partial derivative operator in each direction. This describes how free space polarizations transform in each point group.}
\end{figure}

Next, Fig.~\ref{Group2} provides the Group Theory tables helpful for determining the selection rules through the direct product approach. Figure~\ref{Group2}(a) provides the direct product table for the $C_{6v}$ point group, and Fig.~\ref{Group2}(b) provides the same for the $C_{4v}$ point group. The direct product tables for the lower order point groups are a subset of these. For instance, for $C_{2v}$, Fig.~\ref{Group2}(b) may be used excluding the final row and column. Lastly, Fig.~\ref{Group2}(c) contains the irreducible representations for the partial derivative operators relevant for direct products such as in Eq. (\ref{cc}). This tracks how free space polarizations transform in each lattice type.

\section{Additional Lattices}
\label{B}

\renewcommand{\thefigure}{B~\arabic{figure}}
\setcounter{figure}{0}

\begin{figure}
\includegraphics[width=0.7\columnwidth]{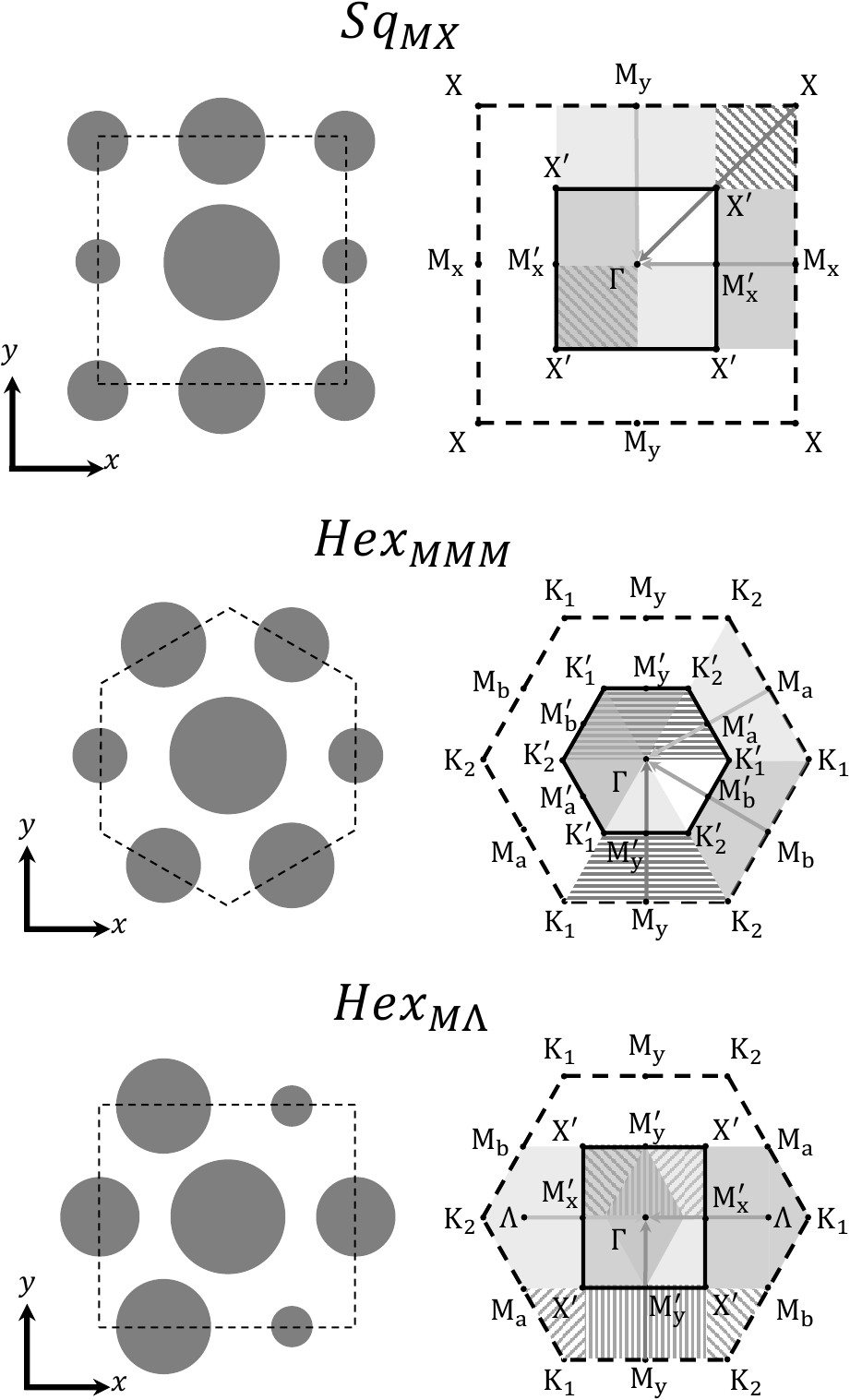}
\caption{\label{B1}Three additional examples of periodically perturbed lattices. Each of these lattices is a "quadromer" lattice, with four atoms per unit cell upon perturbation, and therefore have four times the modes of the unperturbed lattices at the $\Gamma$ point.}
\end{figure}

The six lattices catalogd in Figs.~\ref{fig9} and \ref{fig10} were chosen because they access the six high symmetry modes in the simplest way. For instance, the $Sq_M$ lattice accesses only the $\Gamma$ and $M$ (but not the $X$) modes of a square lattice, while the $Sq_X$ accesses the $\Gamma$ and $X$ (but not the $M$) modes. However, if desired, it is possible to access all three modes in a single lattice by period doubling in both lattice directions. Figure~\ref{B1} shows this lattice, called $Sq_{MX}$, depicting the real space and FBZ. This lattice is a "quadromer", having four atoms per unit cell, and therefore has four times the modes compared to the unperturbed case. In particular, it has the $\Gamma$ and $X$ modes, as well as two copies of the $M$ modes (one from each $M_x$ and $M_y$), which mix at the $\Gamma$ point in a similar way to the $K$ modes of the $Hex_K$ lattice.

Also pictured in Fig.~\ref{B1} are the $Hex_{MMM}$ and $Hex_{M\Lambda}$, both of which are examples of quadromer lattices. $Hex_{MMM}$ is still of the hexagonal lattice family, while $Hex_{M\Lambda}$ (much like $Hex_M$) is rectangular. $Hex_{MMM}$ contains three copies of the $M$ modes, which will mix at the $\Gamma$ point. $Hex_{M\Lambda}$, on the other hand, accesses a unique set of modes at the $\Lambda$ point in the unperturbed FBZ (see the last panel in Fig.~\ref{B1}), which have the point group $C_{s}^d$ (that is, they are either symmetric or anti-symmetric about the $x$ axis). 

The lattices shown in Fig~\ref{B1} are by no means the only additional lattices that may be explored. Instead, they serve as an example of the next few lattices in the infinite list of lattices ordered by number of atoms in the perturbed unit cell. The lattices in this list are generally increasingly complicated, but the same approach outlined in Sec.~\ref{S3} may be applied to determined the selection rules if desired.

\section{Space Group Hierarchy}
\label{C}

\renewcommand{\thefigure}{C~\arabic{figure}}
\setcounter{figure}{0}

A feature of the $Sq_X$ lattice is that there are two points in the unperturbed lattice with $C_{4v}$ about which to apply the dimerizing perturbation. This has the consequence that there are two $p4g$ groups in the catalog of the $Sq_X$ lattice, each with a different high symmetry point in common with the unperturbed lattice (likewise for the two $p4m$ groups). Similarly, there are two equivalent points with at least $C_{2v}$ symmetry that may be chosen while perturbing to a $Sq_M$ lattice (and equivalently, the $Hex_M$ lattice), producing an analogous set of paired perturbations. Note that the $Sq_{\Gamma}$ lattice (and equivalently, the $Hex_{\Gamma}$ lattice) has no such feature. The $Hex_K$ lattice has no such pairing of perturbations because there is only one point in the unperturbed lattice with $C_{6v}$ symmetry.

\begin{figure*}
\includegraphics[width=1.95\columnwidth]{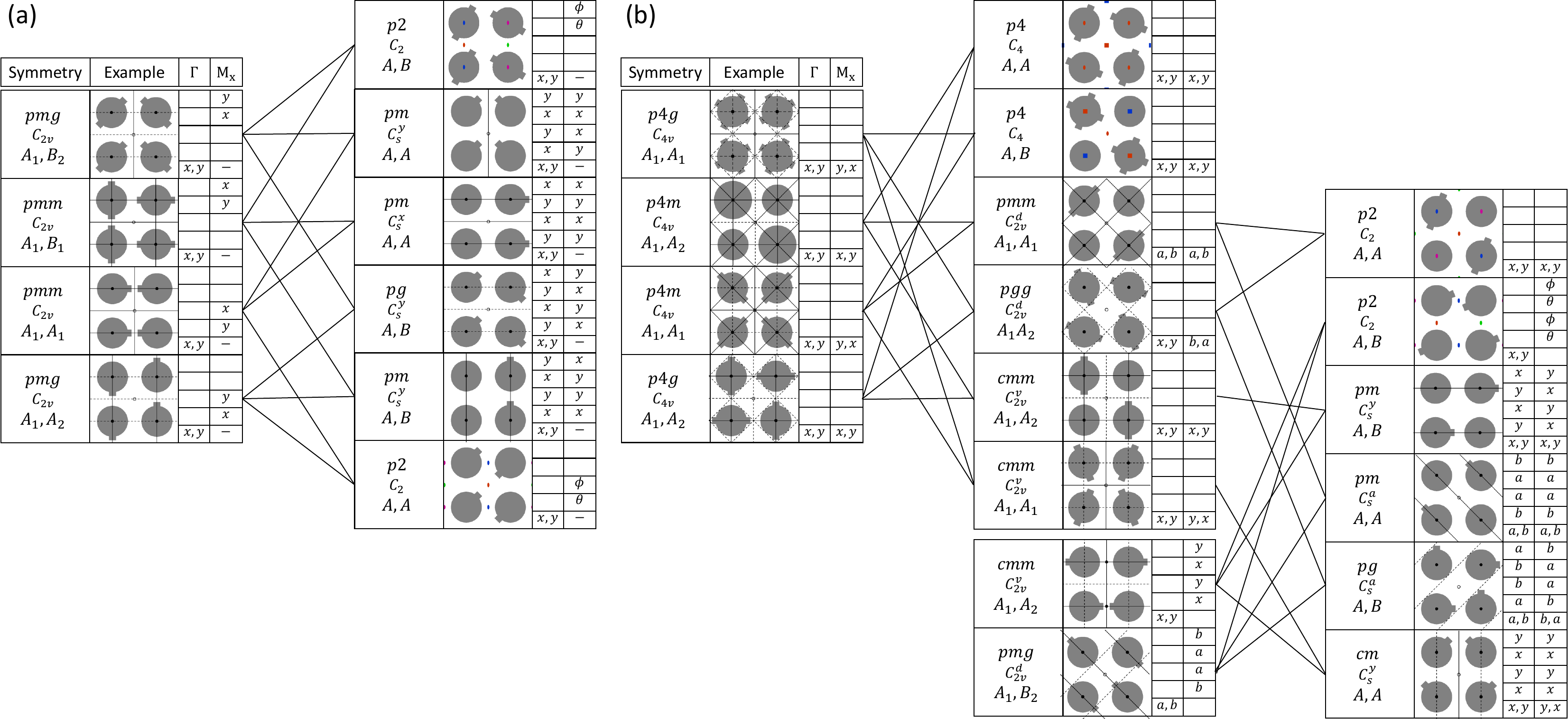}
\caption{\label{figC1} Hierarchy of spacegroups in the (a) $Sq_M$ lattice and (b) $Sq_X$ lattice, showing how lower order space groups are related to higher order space groups.}
\end{figure*}

An interesting feature of the paired sets of perturbation is the relationship of the symmetries those members have in common with each other. By adding the two perturbations together, a space group of lower symmetry is created with only those symmetries the two parent space groups had in common. In the words of crystallography, this process is finding a translationengleiche subgroup in common (and finding the $C_{4v}$ parent space groups is finding klassengleiche subgroups of the unperturbed lattice). This resulting child space group must, naturally, be a member of the catalog. Such a relationship may be studied for all of the space groups in the lattices with this pairing of perturbations. The resulting relationships are seen in Figure~\ref{figC1}, showing complete sets of relationships between higher and lower space groups. Note that for the $Sq_X$ lattice, one of the $cmm$ space groups has its highest symmetry point in common with the $C_{2v}$ symmetry point of the unperturbed lattice. It may therefore not be made by adding higher order space groups. Likewise, it is well known to crystallographers that the $pmg$ space group has no parent (translationengleiche supergroup), and therefore also stands alone. However, both of these ``parent-less'' space groups share symmetries with the other members of the $Sq_X$ lattice, and therefore may be combined to create space groups of lower symmetry. 

We note that while these relationships do not prove that the catalog is exhaustive, the closed, consistent system is highly suggestive that it is. Indeed, in an earlier version of this manuscript, one of the $p4g$ space groups of the $Sq_X$ catalog was omitted, and its existence and selection rules were predicted while attempting constructing Figure~\ref{figC1}(b): a fourth space group with $C_{4v}$ symmetry was needed to produce a consistent hierarchy.

\begin{figure}
\includegraphics[width=1\columnwidth]{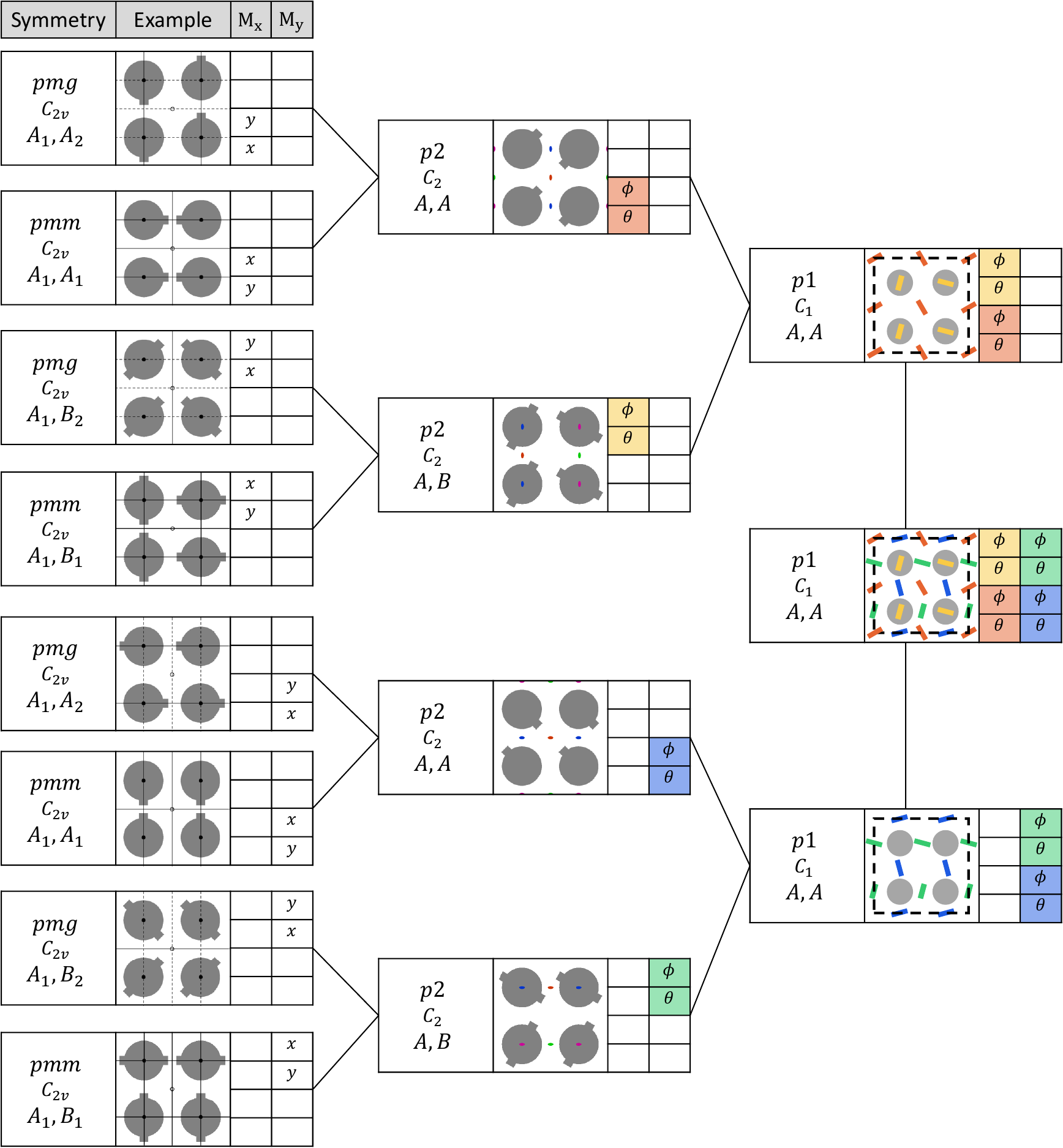}
\caption{\label{figC2}Complete Hierarchy adding eight space groups with $C_{2v}$ point symmetry and ``orthogonal'' selection rules to achieve a final $p1$ space group with eight degree of freedom: the Q-factor and polarization angle of four distinct modes.}
\end{figure}

Finally, for completeness we report in Figure~\ref{figC2} the entire hierarchy enabling control of the Q-factor and polarization angle for four modes simultaneously and independently, used in the accompanying Letter~\cite{overvig_multifunctional_2020}. The resulting lattice is $Sq_{MX}$ (refer to Figure~\ref{B1}) constructed from two $Sq_M$ lattices, one dimerized in the $x$ direction, whose modes are referred to as $M_x$, and the other in the $y$ direction, whose modes are referred to as $M_y$.

\end{document}